\documentclass[review]{elsarticle}
\makeatletter
\def\ps@pprintTitle{%
 \let\@oddhead\@empty
 \let\@evenhead\@empty
 \def\@oddfoot{\centerline{\thepage}}%
 \let\@evenfoot\@oddfoot}
\makeatother

\pdfoutput=1
\usepackage{graphicx}
\usepackage{amsfonts, amsmath, amssymb}
\usepackage{mathabx}
\usepackage{booktabs,siunitx}
\usepackage[svgnames,table]{xcolor}
\usepackage[tableposition=above]{caption}
\usepackage{pifont}
\usepackage{bm}
\usepackage{cancel}
\usepackage{caption}
\usepackage{color}
\usepackage{enumerate}
\usepackage{float}
\usepackage{hyperref}
\usepackage{soul}
\usepackage[english]{babel}
\usepackage[margin=1in]{geometry}
\usepackage{mathtools}
\usepackage{multirow}
\usepackage{setspace}
\usepackage{subfigure}
\usepackage{stmaryrd}
\usepackage{lineno}
\DeclareUnicodeCharacter{00B3}{\textsuperscript{3}}
\usepackage[ruled,linesnumbered]{algorithm2e}
\RestyleAlgo{ruled}
\SetKwComment{Comment}{/* }{ */}

\SetCommentSty{mycommfont}
\DeclareMathAlphabet{\mathpzc}{OT1}{pzc}{m}{it}

\newcommand \dd[2] {\frac{{\rm d} #1}{{\rm d} #2}}

\renewcommand \d[1]{{\rm{d}} #1}
\newcommand \D [2]{\frac{\partial #1}{\partial #2}}

\newcommand \DDD [2]{\frac{{\rm D} #1}{{\rm D} #2}}
\renewcommand{\vec}[1]{\bm{\mathrm{#1}}}
\newcommand{\V}[1]{\bm{\mathrm{#1}}}

\def \div{\nabla \cdot \mbox{}}
\def \grad{\nabla}

\def \x{\vec{x}}
\def \y{\vec{y}}

\def \n{\vec{n}}

\def \u{\vec{u}}

\def \L{\vec{L}}

\def \vD{\vec{D}}

\def \g{\vec{g}}
\def \G{\vec{G}}

\def \Lmu{\vec{L_{\mu}}}
\def \vrho{\vec{\rho}}

\def \Nx{N_x}
\def \Ny{N_y}

\def \Omegal{\Omega_{\text{l}}}
\def \Omegag{\Omega_{\text{g}}}

\def \f{\vec{f}}

\def \half{\frac{1}{2}}
\def \3half{\frac{3}{2}}
\def \5half{\frac{5}{2}}

\def \mul{\mu^{\text{L}}}
\def \mus{\mu^{\text{S}}}
\def \mug{\mu^{\text{G}}}
\def \n{\vec{n}}

\def \nref{n_{\text{ref}}}
\def \ncells{n_{\text{cells}}}
\def \ncycles{n_{\text{cycles}}}

\def \rhol{\rho^{\text{L}}}
\def \rhos{\rho^{\text{S}}}
\def \rhog{\rho^{\text{G}}}
\def \kl{k^{\text{L}}}
\def \ks{k^{\text{S}}}
\def \kg{k^{\text{G}}}

\def \sgn{\textrm{sgn}}

\def \u{\vec{u}}

\def \x{\vec{x}}

\def \div{\nabla \cdot \mbox{}}
\def \grad{\nabla}

\def \dt{\Delta t}
\def \dx{\Delta x}
\def \dy{\Delta y}

\def \rhos{\rho^{\rm S}}
\def \rhol{\rho^{\rm L}}
\def \rhog{\rho^{\rm G}}
\def \cps{C^{\rm S}}
\def \cpl{C^{\rm L}}
\def \cpg{C^{\rm G}}
\def \Ts{T^{\rm S}}
\def \Tl{T^{\rm L}}

\def \ks{\kappa^{\rm S}}
\def \kl{\kappa^{\rm L}}
\def \kg{\kappa^{\rm G}}

\def \ul{u^{\rm L}}
\def \us{u^{\rm S}}

\def \Omegas{\Omega^{\text{S}}}
\def \Omegal{\Omega^{\text{L}}}
\def \Omegam{\Omega^{\text{M}}}
\def \Omegag{\Omega^{\text{G}}}
\def \Omegap{\Omega^{\text{P}}}
\def \alphas{\alpha^{\text{S}}}
\def \alphal{\alpha^{\text{L}}}

\def \mul{\mu^{\text{L}}}
\def \mus{\mu^{\text{S}}}
\def \mug{\mu^{\text{G}}}

\def \Tsol{T^{\rm sol}}
\def \Tliq{T^{\rm liq}}
\def \hsol{h^{\rm sol}}
\def \hliq{h^{\rm liq}}

\newcommand{\upperRomannumeral}[1]{\uppercase\expandafter{\romannumeral#1}}

\newcommand{\REVIEW}[1]{{\color{black}#1}}
\newcommand{\SECONDREVIEW}[1]{{\color{black}#1}}

\begin{document}
\let\today\relax

\begin{frontmatter}
	
\title{A consistent, volume preserving, and adaptive mesh refinement-based framework for modeling non-isothermal gas-liquid-solid flows with phase change}
\author[SDSU]{Ramakrishnan Thirumalaisamy}
\author[SDSU]{Amneet Pal Singh Bhalla\corref{mycorrespondingauthor}}
\ead{asbhalla@sdsu.edu}

\address[SDSU]{Department of Mechanical Engineering, San Diego State University, San Diego, CA}
\cortext[mycorrespondingauthor]{Corresponding author}

\begin{abstract}

This work expands on our recently introduced low Mach enthalpy method~\cite{thirumalaisamy2023lowmach} for simulating the melting and solidification of a phase change material (PCM) alongside (or without) an ambient gas phase. The method captures PCM's volume change (shrinkage or expansion) by accounting for density change-induced flows. We present several improvements to the original work. First, we introduce consistent time integration schemes for the mass, momentum, and enthalpy equations, which enhance the method stability. Demonstrating the effectiveness of this scheme, we show that a system free of external forces and heat sources can conserve its initial mass, momentum, enthalpy, and phase composition. This allows the system to transition from a non-isothermal, non-equilibrium, phase-changing state to an isothermal, equilibrium state without exhibiting unrealistic behavior. Furthermore, we show that the low Mach enthalpy method accurately simulates thermocapillary flows without introducing spurious phase changes. To reduce computational costs, we solve the governing equations on adaptively refined grids. We investigate two cell tagging/untagging criteria and find that a gradient-based approach is more effective. This approach ensures that the moving thin mushy region is always captured at fine grid levels, even when it temporarily falls within a subgrid level. We propose an analytical model to validate advanced computational fluid dynamics (CFD) codes used to simulate metal manufacturing processes (welding, 3D printing). These processes involve a heat source (like a laser) melting metal or its alloy in the presence of an ambient (inert) gas. Traditionally, studies relied on artificially manipulating material properties to match complex experiments for validation purposes. Leveraging the analytical solution to the Stefan problem with a density jump, this model offers a straightforward approach to validating multiphysics simulations involving heat sources and phase change phenomena in three-phase flows. Lastly, we demonstrate the practical utility of the method in modeling porosity defects (gas bubble trapping) during metal solidification. A field extension technique is used to accurately apply surface tension forces in a three-phase flow situation. This is where part of the bubble surface is trapped within the (moving) solidification front. 
    
\end{abstract}

\begin{keyword}
\emph{Stefan problem} \sep \emph{volume shrinkage/expansion} \sep \emph{low Mach formulation} \sep \emph{metal manufacturing} \sep \emph{high density ratio flows} 
\end{keyword}

\end{frontmatter}

\section{Introduction}
  
Metal manufacturing relies heavily on phase change phenomena, particularly melting and solidification, to create parts through processes such as welding, casting, and additive manufacturing (AM)/3D printing~\cite{katayama2013handbook,blakey2021metal}. Many factors, such as laser power, scan speed, and inert gas flow rate in processes such as selective laser melting (of metals), can affect the quality of the final product~\cite{spierings2018influence}. Understanding how these parameters influence defects like porosity, keyhole formation, and spattering is crucial for manufacturing good quality parts~\cite{abouelnour2022situ}. Numerical modeling offers valuable insights into these difficult-to-observe defects in-situ and optimizes manufacturing.

While significant progress has been made in modeling phase change, it remains an active research area. Among various methods, fixed-grid CFD techniques are popular due to their ability to capture movement and topological changes to the phase change front~\cite{dutil2011review}. These techniques are further categorized as sharp-interface~\cite{ye2001fixed, gibou2007level, sato2013sharp} and diffuse-interface methods~\cite{voller1987fixed, voller1991eral, boettinger2002phase, beckermann1999modeling,huang2022consistent}. Sharp-interface methods enforce strict jump conditions across the phase boundary, offering high accuracy but becoming computationally expensive or perhaps even infeasible for practical scenarios involving more than two phases; for example, solid, liquid and vapor phases of metal and an additional ambient inert gas phase. Resolving four distinct phases and their interactions becomes extremely challenging with sharp-interface techniques. Additionally, simulating manufacturing processes over a long duration (e.g., till equilibrium) can involve phases appearing or disappearing entirely, which can cause issues with imposing jump conditions (which are expressed as spatial gradients) in sharp-interface formulations. Diffuse-interface methods, on the other hand, can handle these complexities relatively easily. The enthalpy method (EM) is a popular diffuse-interface approach to modeling phase change, and this work proposes advancements to improve its stability, efficiency, and ability to handle additional physics like thermocapillary flows and three-phase surface tension models.



The fixed-grid EM was introduced by Voller and colleagues~\cite{voller1987fixed,voller1991eral}  for modeling the melting and solidification of phase change materials (PCMs). In contrast to sharp interface techniques, EM models the phase transition process across a finite ``mushy" region, which regularizes jumps in various quantities (heat flux, velocity, kinetic energy, etc.) across the phase change boundary. Many research studies have used the EM to model various phase change applications~\cite{lin2020conservative, panwisawas2017keyhole, galione2015fixed, faden2019optimum}. FLOW-3D \cite{flow3d}, the most widely used commercial software for modeling industrial welding and additive manufacturing processes, also implements Voller's standard enthalpy method. In spite of its widespread adoption for modeling melting and solidification problems, EM has not evolved much since its introduction. The standard EM also does not account for the fluid motion caused by density differences between liquid and solid phases, and consequently, it does not capture volume changes caused by melting and solidification. Recently, we proposed a low Mach enthalpy method \cite{thirumalaisamy2023lowmach} to address this limitation. Our method allows for changes in material volume within the narrow mushy region by incorporating velocity constraints derived from the equation of state and conservation of mass equation. Additionally, our approach enables the coupling of a solid-liquid PCM with a gas phase to simulate PCM's free surface dynamics.  

This work introduces several improvements that enhance the low Mach enthalpy method for modeling melting and solidification problems. First, we improve the stability of the model by using consistent time integrators for mass-momentum-enthalpy equations. In addition, we incorporate adaptive mesh refinement (AMR) to simulate metal manufacturing processes, which are computationally very expensive to simulate with uniform grids, especially in three dimensions. We propose a gradient-based tagging method that ensures fine grids capture narrow mushy regions, even as the mushy regions disappear (i.e., temporarily fall within subgrid levels) during simulation. The low Mach method is also used to simulate thermocapillary flows with variable surface tension, since temperature-dependent surface tension plays a key role in AM processes~\cite{cook2020simulation}. We provide an analytical model to validate CFD codes with a heat source (like a laser) in the presence of a gas phase. This approach offers an alternative to studies that artificially manipulate material properties in the numerical simulation to match complex experiments \cite{saldi2012marangoni, saldi2013effect, ebrahimi2020numerical, pitscheneder1996role} for validation purposes. The method is also applied to model porosity defects (gas bubble trapping during metal solidification) using adaptively refined grids. We adopt a partial differential equation (PDE)-based field extension technique that enables accurate application of surface tension force in a three-phase flow situation, in which some part of the gas bubble is trapped within the (moving) solidification front while the remainder resists deformation (due to surface tension) as a result of fluid flow. In the literature, phase change simulations rarely verify whether a numerical scheme can transition from a non-equilibrium to an equilibrium state reliably. Most simulations are run until equilibrium in the current work. This is to ensure that the improved low Mach EM can handle complete phase appearance and disappearance in a stable manner.  


\section{Mathematical formulation} \label{sec_formulation}
\SECONDREVIEW{In this section, we succinctly write the low Mach enthalpy equations derived in our previous work~\cite{thirumalaisamy2023lowmach} for reader's convenience. We also present a thermodynamic justification for the low Mach enthalpy model, a three-phase surface tension model, and a heat source term in the enthalpy equation, which were not discussed in our previous work~\cite{thirumalaisamy2023lowmach}.}
\subsection{PCM-gas interface tracking}

\begin{figure}[]
\centering
\includegraphics[width=0.4\linewidth]{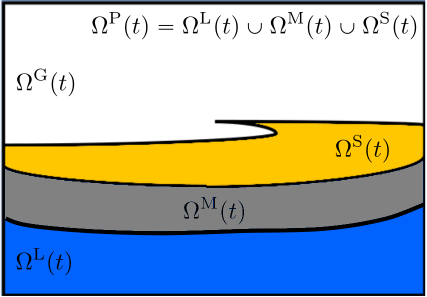}
\caption{\REVIEW{A schematic depicting three phases: liquid (represented in blue), solid (in yellow), and gas (in white). The gray region represents the mushy region $\Omegam(t)$. This is a very thin region, but it has been exaggerated for illustration purposes. A Heaviside function $H(\REVIEW{\Phi})$ is used to distinguish between the PCM $(\Omegap(t) = \Omegal(t) \cup \Omegam(t) \cup \Omegas(t))$ and gas $\Omegag(t)$ regions. The liquid fraction variable $\varphi$ tracks the liquid-solid interface. The conventions adopted in this study are as follows: $H=1$ within the PCM, $H=0$ within the gas, and $0<H<1$ within the interface cells. $\varphi=1$ within the liquid phase, $\varphi=0$ within the solid phase, and $0<\varphi<1$ within the mushy region.}}
\label{fig_schematic}
\end{figure}

The static computational domain $\Omega$ consists of a time varying PCM region $\Omegap(t)$ and a time varying gas region $\Omegag(t)$, such that $\Omega = \Omegap(t) \cup \Omegag(t) \; \forall t$; see Fig.~\ref{fig_schematic}.  The PCM-gas interface $\Gamma(t)$ is tracked using the level set method. In this approach, the interface  is defined as the zero contour of the level set function $\REVIEW{\Phi}(\x,t)$ 
\begin{equation}
\Gamma(t) = \{\x\in\Omega\, | \,\REVIEW{\Phi}(\x,t)=0\}.
\end{equation}
The level set function $\REVIEW{\Phi}(\mathbf{x},t)$ is taken to be a signed distance function, and defined as
\begin{align}
\REVIEW{\Phi}(\x,t) = \begin{cases}
 > 0 ,&  \text{PCM},\\ 
 < 0 ,&  \text{gas}, \\ 
 =0,&  \text{PCM-gas interface}.
\end{cases}
\label{eq_ls_def}
\end{align}
$\REVIEW{\Phi}$ is advected with a non-divergence free velocity $\u(\x,t)$ according to
\begin{equation}
\DDD{\REVIEW{\Phi}}{t} = \D{\REVIEW{\Phi}}{t}+\u\cdot\grad{\REVIEW{\Phi}}=0. 
\label{eq_level_set}
\end{equation}
Under the action of advection, $\REVIEW{\Phi}$ looses its the signed distance property. The signed distance property is lost at the continuous level (due to advection), and it should not be viewed as a purely numerical problem~\cite{khedkar2024preventing}.  
\REVIEW{A reinitialization procedure proposed by Sussman et al.~\cite{Sussman1994} is employed to retain the signed distance property of $\REVIEW{\Phi}(\x,t)$. This is achieved by computing the steady-state
solution to the Hamilton-Jacobi equation
\begin{subequations} 
\begin{alignat}{2}
&\D{\Phi}{\tau} + \sgn\left(\widetilde{\Phi}\right)\left(\|\grad \Phi \| - 1\right) = 0, \label{eq_eikonal} \\
& \Phi(\x, \tau = 0) = \widetilde{\Phi}(\x). \label{eq_eikonal_init}
\end{alignat}
\end{subequations} 

At the end of a physical time step, Eq.~\eqref{eq_eikonal} is evolved in pseudo-time $\tau$, which, at steady state, produces
a signed distance function satisfying the Eikonal equation $\|\grad \Phi \|  = 1$. Here, $\sgn$ denotes the sign of
$\widetilde{\Phi}$, which is either $1$, $-1$, or $0$. To mitigate mass loss, we employ the subcell-fix method proposed by Min~\cite{min2010reinitializing} and enforce an immobility condition near the zero level set as described by Son~\cite{son2005level}. For more details about the reinitialization procedure used in this work, please refer to our previous papers~\cite{nangia2019robust, thirumalaisamy2023pre}.} The level set field is reinitialized at each time step.  A Heaviside function $H(\x, t)$ is defined using the level set function $\REVIEW{\Phi}(\x,t)$ to distinguish the PCM  and gas regions. 
We use a smooth Heaviside function to smoothly transition material properties from the gas to the PCM domain. Its functional form reads as 
\begin{align}
H &= \begin{cases}
 0,& \REVIEW{\Phi}(\x)<-n_\text{cells} \, \Delta,\\ 
 \frac{1}{2}\left[1+\frac{1}{n_\text{cells} \, \Delta} \; \REVIEW{\Phi}(\x) +  
     \frac{1}{\pi}\text{sin}\left(\frac{\pi}{n_\text{cells} \, \Delta} \; \REVIEW{\Phi}(\x)\right)\right],& |\REVIEW{\Phi}(\x)| \le n_\text{cells} \, \Delta,\\ 
 1,&\text{otherwise}.
\end{cases}
\label{eq_smooth_H}
\end{align}
Here, $n_\text{cells} \Delta$ is the interfacial region's width on either side of the gas-PCM interface over which the Heaviside function is smoothed, and $n_\text{cells} \in \mathbb{R}$ and $\Delta$ represent the number of grid cells and cell size, respectively. Unless otherwise specified, $\ncells = 2$ is used in all simulations in this work. 

\subsection{Enthalpy equation}
Within the PCM region $\Omegap(t)$, three distinct phases can coexist: solid $\Omegas(t)$, liquid $\Omegal(t)$, and mush $\Omegam(t)$. A liquid fraction variable $\varphi(\x,t)$ is used to track the composition throughout the PCM: $\varphi \equiv 1~~\forall~\x~\in \Omegal(t)$,  $\varphi \equiv 0~~\forall~\x~\in \Omegas(t)$, and $ 0 < \varphi < 1~~\forall~\x~\in \Omegam(t)$. The evolution of $\varphi$ is determined by the enthalpy equation, which is expressed as
\begin{equation}
\D{\left(\rho h\right)}{t}+\div{\left(\rho \u h\right)}= \div\left({\kappa\grad{T}}\right) + Q_{\rm src}.
\label{eq_enthalpy}
\end{equation}
Here, $h(\x,t)$ is the specific enthalpy, $T(\x, t)$ represents temperature, $\kappa (\x, t)$ is the thermal conductivity, and $Q_{\rm src}$ denotes a thermal source term (e.g., laser heating). The enthalpy Eq.~\eqref{eq_enthalpy} is solved in the entire domain, i.e., in both gas and PCM regions. The specific enthalpy $h$ of solid, liquid, and mush are defined in terms of their temperature $T$ as
\begin{align}
h = \begin{cases}
 \cps (T - T_r) ,&  T<\Tsol,\\ 
 \bar{C}(T - \Tsol) + \hsol + \varphi \frac{\displaystyle \rhol}{\displaystyle \rho}L,&\Tsol \le T \le \Tliq, \\ 
 \cpl(T-\Tliq)+\hliq,& T> \Tliq ,
\end{cases}
\label{eq_h_pcm}
\end{align}
and of the gas as
\begin{equation}
h = \cpg (T - T_r) .
\label{eq_h_gas}
\end{equation} 
\REVIEW{Here, $\cpl$, $\cps$ and $\cpg$ are the specific heats of liquid, solid and gas, respectively.} The gas and PCM domains are distinguished by the $\REVIEW{\Phi}(\x,t) = 0$ or $H(\x,t) = 0.5$ contours. Eq.~\eqref{eq_h_pcm} introduces two key temperatures: liquidus temperature $\Tliq$ where solidification begins and solidus temperature $\Tsol$ where it completes. For pure materials, there is a single, well-defined temperature $T_m$ at which phase change occurs. However, alloys and glassy materials exhibit a more gradual phase change process that happens over a range of temperature $\Delta T$ = $\Tliq - \Tsol$. The enthalpy method~\cite{thirumalaisamy2023lowmach} employed in this work takes the latter approach and models the melting/solidification process over an extended temperature range. The solidus and liquidus enthalpies are defined as $\hsol = \cps (\Tsol - T_r)$ and $\hliq = \bar{C}(\Tliq-\Tsol)+\hsol+L$, respectively. Here, $\bar{C}=\frac{\cps+\cpl}{2}$ is the specific heat of the mushy region. It is taken as the average of liquid $\cpl$ and solid $\cps$ specific heats. The latent heat of melting/fusion is denoted $L$, and the specific heat of gas is denoted $\cpg$.  

Eqs.~(\ref{eq_h_pcm}) and (\ref{eq_h_gas}) imply that solid and gas enthalpies are zero at $T = T_r$. The numerical solution is not affected by this arbitrary choice of reference temperature $T_r$, and in the numerical simulations we set $T_r = T_m$. We use a mixture model  to express density and specific enthalpy in terms of liquid fraction in the mushy region 
\begin{subequations}
\begin{alignat}{2}
\rho &= \varphi \rhol + (1-\varphi) \rhos   \label{eq_rho_mixture}, \\
\rho h &= \varphi \rhol \hliq + (1-\varphi) \rhos \hsol.   \label{eq_h_mixture}
\end{alignat}
\end{subequations}
Here, $\rhos$ and $\rhol$ are the densities of  solid and liquid phases of the PCM. Substituting $h$ from Eq.~\eqref{eq_h_pcm} and $\rho$ from Eq.~\eqref{eq_rho_mixture} into Eq.~\eqref{eq_h_mixture}, we obtain a $\varphi$-$T$ relation for the mushy region 
\begin{equation}
\varphi = \frac{\displaystyle \rho}{\displaystyle \rhol}\frac{\displaystyle T-\Tsol}{\displaystyle \Tliq-\Tsol}.
\label{eq_varphi_mixture}
\end{equation}
Knowing $\varphi$ in terms of $T$ (Eq.~\eqref{eq_varphi_mixture}) allows us to invert $h$-$T$ relations. The  temperature in the solid-liquid-mushy region
\begin{align}
T = \begin{cases}
 \frac{\displaystyle h}{\displaystyle \cps} + T_r, & h<\hsol,\\ 
  \Tsol + \frac{\displaystyle h-\hsol}{\displaystyle \hliq-\hsol}(\Tliq-\Tsol),&\hsol \le h \le \hliq, \\ 
 \Tliq + \frac{\displaystyle h -\hliq}{\displaystyle \cpl},& h > \hliq,
\end{cases}
\label{eq_T_pcm}
\end{align} 
and in the gas region
\begin{equation}
T =  \frac{\displaystyle h}{\displaystyle \cpg} + T_r
\label{eq_T_gas}
\end{equation}
can be written in terms of $h$. These $T$-$h$ relations are used in Newton's iterations to solve the nonlinear enthalpy Eq.~\eqref{eq_enthalpy}. We will discuss this in more detail later. Similarly, substituting $\rho$ from Eq.~\eqref{eq_rho_mixture} into Eq.~\eqref{eq_h_mixture},  we get  a $\varphi$-$h$ relation 
\begin{align}
\varphi = \begin{cases}
 0,& h < \hsol,\\ 
  \frac{\displaystyle \rhos(\hsol-h)}{\displaystyle h(\rhol-\rhos)-\rhol \hliq + \rhos \hsol},&\hsol \le h \le \hliq, \\ 
 1,& h > \hliq.
\end{cases}
\label{eq_liquid_fraction}
\end{align}
Although arbitrary, $\varphi$ in the gas region is defined to be zero.

\subsubsection{Mixture model}

Eq.~\eqref{eq_enthalpy} requires specifying thermophysical properties like $\rho$ and $\kappa$ (and later viscosity $\mu$) across the entire computational domain. Denoting a generic thermophysical property with $\beta$, it can be expressed using a mixture model of the form 
\begin{subequations} 
\begin{alignat}{2}
& \beta = \beta^{\rm{G}}+( \beta^{\rm{S}}- \beta^{\rm{G}})H+( \beta^{\rm{L}}- \beta^{\rm{S}})H\varphi.  \label{eq_material_properties}  \\
& \rho= \rhog+( \rhos- \rhog)H+( \rhol- \rhos)H\varphi.  \label{eq_EOS}
\end{alignat}
\end{subequations} 
When $\beta = \rho$, we get the equation of state (EOS) written in Eq.~\eqref{eq_EOS}. The EOS  imposes a kinematic constraint on the velocity field as shown in the next section. Note that Eq.~\eqref{eq_rho_mixture} corresponds to the EOS, which is obtained by substituting $H(\Omegap) \equiv 1$ in Eq.~\eqref{eq_EOS}.

\subsection{Low Mach Navier-Stokes equations}
We use the one-fluid formulation to represent the motion of solid, liquid, mush, and gas within the  
computational domain $\Omega$. The momentum of the multiphase system is governed by Navier-Stokes equations, which read as 
\begin{subequations} 
\begin{alignat}{2}
\D{\left(\rho \u\right)}{t}+\div{\left(\rho \u \otimes \u\right)} &= -\grad{p}+\div\left[{\mu\left(\grad{\u}+\grad{\u}^T\right)}\right] 
+\rho \g - A_d \u + \f_{\rm st}, \label{eq_momentum} \\
\div \u & = -\frac{1}{\rho}\DDD{\rho}{t}. \label{eq_continuity}
\end{alignat}
\end{subequations}
Here, $\u(\x, t)$ denotes the Eulerian velocity, $p(\x, t)$ represents the Eulerian pressure, and $\rho(x, t)$ and $\mu(\x,t)$ are spatially and temporally varying density and viscosity fields, respectively.  $\rho(\x,t)\, \g$ in the momentum Eq.~\eqref{eq_momentum} represents the gravitational body force, with $\g$ denoting the acceleration due to gravity.  \REVIEW{The drag force term $-A_d \u$, commonly referred to as the Carman-Kozeny force~\cite{voller1991eral} is used to retard any flow in the solid region.} $\f_{\rm st}(\x, t)$ represents the surface tension force at the gas-liquid interface, which will be discussed later.  The continuity Eq.~\eqref{eq_continuity} provides a constraint on the velocity field and is directly derived from the conservation of mass equation:
\begin{subequations}
\begin{alignat}{2}
& \D{\rho}{t}+\div{\left(\rho \u\right)}=0, \label{eq_mass_blance} \\
 \hookrightarrow & \D{\rho}{t}+\rho \div{\u} + \u \cdot \grad{\rho}=0, \\
\hookrightarrow & \div{\u}=-\frac{1}{\rho}\left(\D{\rho}{t}+\u \cdot \grad{\rho}\right) = -\frac{1}{\rho}\DDD{\rho}{t}.
\end{alignat}
\end{subequations}

The EOS written in Eq.~\eqref{eq_EOS} allows us to express the material derivative of density in the PCM domain in terms of material derivatives of liquid fraction and enthalpy 
\begin{subequations} 
\begin{align}
\div \u  = &-\frac{1}{\rho}\DDD{\rho}{t} = \frac{(\rhos-\rhol)}{\rho} H \DDD{\varphi}{t}, \\   
\DDD{\varphi}{t}  = &\begin{cases}
 & 0,\hspace{16.6em} h<\hsol,\\ 
 & \frac{\displaystyle -\rhos \rhol (\hsol-\hliq)}{\displaystyle (h(\rhol- \rhos)- \rhol \hliq + \rhos \hsol)^2}\displaystyle \DDD{h}{t}, \qquad \hsol \le h \le \hliq, \\ 
 & 0,  \hspace{16.6em} h> \hliq ,
\end{cases}
\label{eq_dvarphi_dt}  \\
\DDD{h}{t}  = & \frac{1}{\rho} \left( \div\left({\kappa\grad{T}}\right) + Q_{\rm src} \right). \label{eq_dH_dt}
\end{align}
\end{subequations}
Eqs.~\eqref{eq_dvarphi_dt} and~\eqref{eq_dH_dt} are derived from Eqs.~\eqref{eq_liquid_fraction} and~\eqref{eq_enthalpy}, respectively. The continuity Eq.~\eqref{eq_continuity} in the entire computational domain reads as
\begin{align}
& \div{\u} = \begin{cases} 
0, & H < 0.5 \; (\text {i.e., in the gas phase}), \\
0,& h <\hsol,\\ 
-\frac{ \rhos \rhol}{\rho^2}(\rhol-\rhos)H \frac{\displaystyle (\hliq-\hsol)}{\displaystyle \left(h(\rhol- \rhos)- \rhol \hliq +\rhos \hsol \right)^2}\displaystyle \left(\div {\kappa\grad{T}}  + Q_{\rm src} \right),&\hsol \le h \le \hliq, \\ 
0,& h >\hliq.\\ 
\end{cases} \label{eq_lowmach_discretized}
\end{align}

Eq.~\eqref{eq_lowmach_discretized} suggests that velocity $\u$ is divergence-free in the gas, solid and liquid regions and non-divergence-free in the mushy region. In other words, density-induced volume changes occur only in the mush. Eq.~\eqref{eq_lowmach_discretized} corresponds to the low Mach formulation of the compressible Navier-Stokes equations. The scaling analysis of the next section justifies the low Mach formulation of the compressible Navier-Stokes equations.       

\subsection{Thermodynamics of the phase change problem}

In the general case of a compressible fluid/material, density is expressed in terms of pressure and temperature as $\rho = \rho(p,T)$. The material derivative of density reads as
\begin{equation}
\frac{1}{\rho}\DDD{\rho}{t} = \alpha \DDD{p}{t} - \beta \DDD{T}{t}. \label{eq_thermod_drhodt}
\end{equation}
Here, $\displaystyle{\alpha(p,T) = \frac{1}{\rho}\D{\rho}{p}\bigg|_T}$ and $\displaystyle{\beta(p,T) = -\frac{1}{\rho}\D{\rho}{T}\bigg|_p}$ are the isothermal compressibility and bulk expansion coefficients, respectively. While $\alpha$ and $\beta$ represent the material's inherent thermodynamic response, $\DDD{p}{t}$ and $\DDD{T}{t}$ represent the flow field effects. To identify dominant terms during phase change, we re-write Eq.~\eqref{eq_thermod_drhodt} in dimensionless form. This involves scaling all variables based on characteristic values. Temperature, for instance, is scaled as $\displaystyle{T^* = \frac{T - T_i}{T_w - T_i}}$, in which $T_i$ denotes the initial temperature of the PCM and $T_w$ is the wall temperature which is sufficiently different from $T_i$ to induce phase change. Similar scaling applies to density, time, pressure, and other relevant quantities: \[\rho^* = \frac{\rho}{\rho_0}~,t^* = \frac{tv_0}{l},~p^* = \frac{p}{\rho_0v_0^2}~,\alpha^* = \frac{\alpha}{\alpha_0},~\beta^* = \frac{\beta}{\beta_0},~Ma = \frac{v_o}{C_s}.\] Here, $Ma$ denotes the Mach number, $C_s = \sqrt{\D{p}{\rho}\big|_s}$ is the speed of sound, and $l$, $\rho_0$, $v_0$, $\alpha_0$ and $\beta_0$ are the reference values of length, density, velocity and thermodynamic functions, respectively.
Eq.~\eqref{eq_thermod_drhodt} when non-dimensionalized reads as
\begin{equation}
\frac{1}{\rho^*}\DDD{\rho^*}{t^*} = \gamma Ma^2 \alpha^*\DDD{p^*}{t^*} - B \beta^* \frac{T_w - T_i}{T_i}\DDD{T^*}{t^*},  \label{eq_thermod_drhodt_nd}
\end{equation}
in which $\gamma = C_p/C_v = \rho_0\alpha_0C_s^2$ and $B = T_i \beta_0$ are non-dimensional (thermodynamic) parameters. 

Our model treats the bulk solid and liquid regions of the phase change material as incompressible. This means the speed of sound within these regions is essentially infinite. Since the mushy region between them is extremely thin, on the order of a few atomic/molecular diameters, its sound speed is also expected to be close to infinity. With these assumptions, the Mach number (ratio of flow velocity to sound speed) becomes negligible. As a consequence, Eq.~\eqref{eq_thermod_drhodt_nd} (or its dimensional form, Eq.~\eqref{eq_thermod_drhodt}) suggests that density changes are primarily driven by temperature variations, not pressure fluctuations
\begin{equation}
\frac{1}{\rho}\DDD{\rho}{t} \approx  - \beta \DDD{T}{t}. \label{eq_lm_drhodt}
\end{equation} 
Thus, under the low Mach assumption, pressure $p$ is treated as a mechanical variable rather than a thermodynamic one. In other words, pressure acts as a Lagrange multiplier to satisfy kinematic constraints on velocity and does not affect density. Note that Eq.~\eqref{eq_lowmach_discretized} is a different way of expressing Eq.~\eqref{eq_lm_drhodt} as $\DDD{T}{t}$ can be expressed in terms of enthalpy: $\DDD{T}{t} = \D{T}{h}\DDD{h}{t}$. Moreover the bulk expansion coefficient of the mushy region corresponding to the mixture model Eq.~\eqref{eq_EOS} is $\beta = \frac{\rhos - \rhol}{\rho}\D{\varphi}{T}$. $\beta$ is zero for solid, liquid and gas phases.          

\subsection{Surface tension force}

A continuum surface tension model proposed by Brackbill et al. \cite{brackbill1992continuum} is used to account for surface tension $\f_\text{st}$ along the liquid-gas interface. The surface tension forces in flows with large temperature and concentration gradients are not uniform. Temperature-dependent surface tension causes additional thermocapillary flows. We assume that the surface tension coefficients $\sigma$ are temperature-dependent in this work. This assumption leads to a continuum surface tension force of the form
\begin{subequations}
\begin{alignat}{2}
\f_{\rm st} &= \sigma(T) \mathcal{C}(\REVIEW{\Phi}) \delta \n + \nabla_{||} \sigma(T) \, \delta,	\label{eq_fst_a}\\ 
\hookrightarrow \V{f}_{\rm st} &= \sigma(T) \mathcal{C}(\REVIEW{\Phi}) \delta \n +  \left [\grad{\sigma(T)} -  \left( \grad{\sigma(T)} \cdot \n \right)\n \right] \, \delta. \label{eq_fst_b}
\end{alignat}
\end{subequations}
Here, $\mathcal{C}(\REVIEW{\Phi}) = -\div \n = -\div \left( \frac{\grad{\REVIEW{\Phi}}}{|\grad{\REVIEW{\Phi}}|} \right)$ is the curvature of the interface computed from the level set function $\REVIEW{\Phi}$, and $\delta$ is the  smoothed/mollified delta function. Implementing Eq.~\eqref{eq_fst_b} as is leads to spurious/parasitic velocity currents near the interface. Francois et al.~\cite{francois2006balanced} proposed a well-balanced scheme in which $\delta$ and $\delta \, \n$ are obtained from the discrete gradient of a mollified Heaviside function $\widetilde{B}$ as $\delta = |\grad {\widetilde{B}}|$ and $\delta \, \n = \grad \widetilde{B}$. \REVIEW{We use Peskin’s four-point regularized delta function~\cite{peskin2002immersed} to mollify the Heaviside function as 
\begin{equation} 
\widetilde{B}(\x,t) = \int_\Omega H(\y,t) \delta(\y - \x) \d \y, 
\end{equation} 
in which $\delta(\y - \x)$ represents the four-point regularized delta function.}

Furthermore, the spatial gradient of the surface tension coefficient can be expressed as $\grad{\sigma} = \frac{\rm d \sigma}{\rm d T} \, \grad{T}$. With these substitutions, Eq.~\eqref{eq_fst_b} becomes
\begin{subequations} \label{eq_marangoni}
\begin{alignat}{2}
 \V{f}_{\rm st} & = \sigma(T) \mathcal{C}(\REVIEW{\Phi}) \grad {\widetilde{B}} +  \grad{\sigma(T)} |\grad {\widetilde{B}}| -  \left( \grad{\sigma(T)} \cdot \frac{\grad{\REVIEW{\Phi}}}{|\grad{\REVIEW{\Phi}}|} \right) \, \grad {\widetilde{B}},\label{eq_fst_c}	\\ 
\hookrightarrow \V{f}_{\rm st} & = \sigma(T) \mathcal{C}(\REVIEW{\Phi}) \grad {\widetilde{B}} +  \frac{\rm d \sigma}{\rm d T} \grad{T} |\grad {\widetilde{B}}| -  \left( \grad{T} \cdot \frac{\grad{\REVIEW{\Phi}}}{|\grad{\REVIEW{\Phi}}|} \right) \, \frac{\rm d \sigma}{\rm d T}  \grad {\widetilde{B}}.  \label{eq_fst_d}
\end{alignat}
\end{subequations}
$ \frac{\rm d \sigma}{\rm d T}$ is commonly referred to as the Marangoni surface tension coefficient.

\begin{figure}[]
\centering
\includegraphics[width=0.9\linewidth]{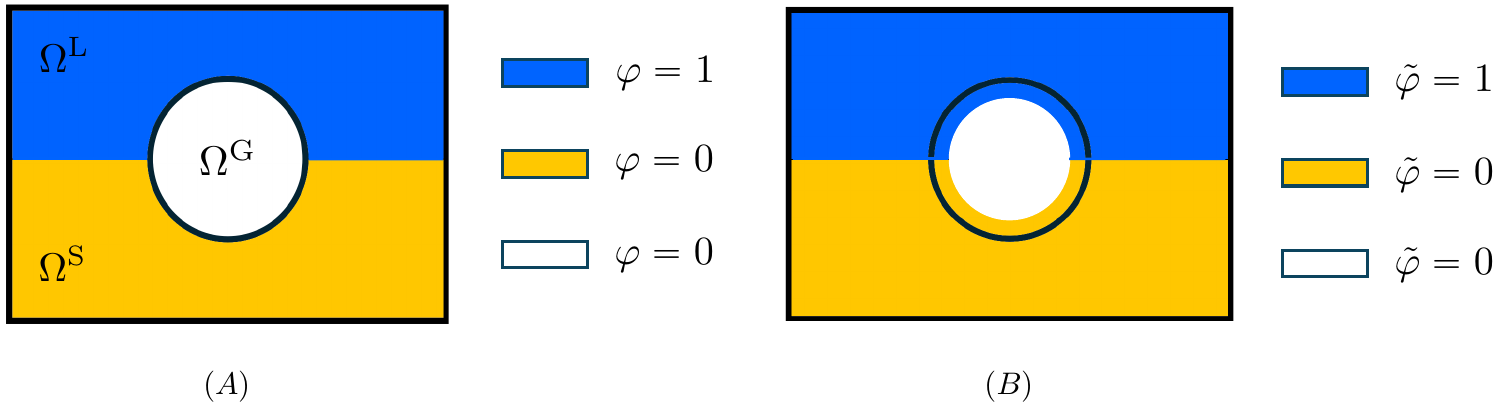}
\caption{Illustration of (A) liquid fraction variable $\varphi$ used in setting  thermophysical properties; and (B) extended liquid fraction variable $\widetilde{\varphi}$ used in computing three-phase surface tension force $\V{f}_{\rm st}^{\rm 3-phase}$.}
\label{fig_lf_extrap_schematic}
\end{figure}

\begin{figure}[]
\centering
\includegraphics[width=0.9\linewidth]{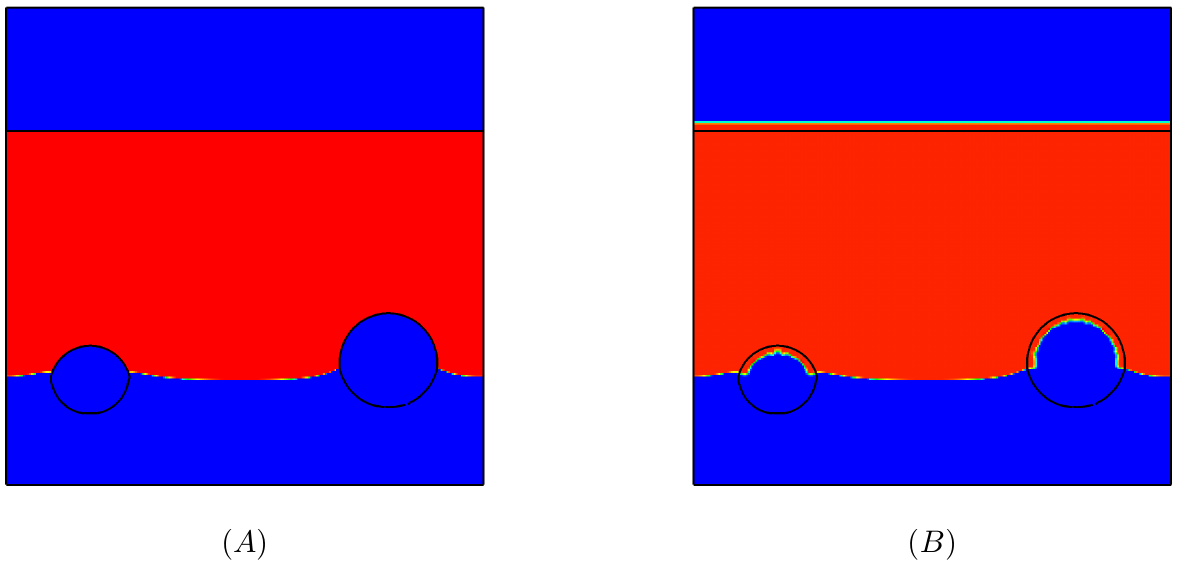}
\caption{Illustration of (A)~liquid fraction ($\varphi$) and \REVIEW{(B)~} extended liquid fraction ($\widetilde{\varphi}$) variables at a representative time for the case of gas bubble trapping during metal solidification considered in Sec.~\ref{sec_bubble_trap}.}
\label{fig_bubble_trap_schematic}
\end{figure}

The form of the surface tension force written in Eqs.~\eqref{eq_marangoni} is appropriate for a two-phase flow (liquid and gas) model. In three-phase flows involving liquid-solid-gas interfaces, the surface tension force form needs to be adjusted. For simulating metal manufacturing applications, gas-solid and liquid-solid surface tensions can be ignored. With this assumption, the overall three-phase dynamic can be fairly well captured, except for achieving the correct contact angle at the material triple points. In highly non-equilibrium situations like welding and metal AM, dynamic contact angle conditions are rarely known, and not imposing one is a reasonable compromise. In light of these assumptions, our three-phase surface tension force model ignores contributions from solid-gas and solid-liquid interfaces. One approach to achieving this (\REVIEW{ignoring surface tension of the solid phase}) is to multiply the two-phase surface tension force $\V{f}_{\rm st}$ by the liquid fraction variable $\varphi$, since $\varphi = 0$ within the solid domain. Therefore, a possible form of the three-phase flow surface tension force reads as
\begin{equation*}
\V{f}_{\rm st}^{\rm 3-phase} = \V{f}_{\rm st} \varphi. \label{eq_st_3phase}
\end{equation*} 

Although the above form of $\V{f}_{\rm st}^{\rm 3-phase}$ appears straightforward at first glance, the continuum surface tension model of Brackbill et al. introduces an additional complexity---the surface tension force acts on both sides of the liquid-gas interface. Within the gas region $\Omegag$, $\varphi$ is not defined as it is not needed. In our simulations, we set $\varphi = 0$ inside $\Omegag$, which is an arbitrary choice. If $\varphi = 0 \; \forall\; \x \in \Omegag$ is used to define $\V{f}_{\rm st}^{\rm 3-phase}$, then the diffuse form of surface tension force acts only within one-half of the interfacial region on the PCM side. This is clearly incorrect. To resolve this situation $\varphi$ needs to be extended into the gas region. The need for $\varphi$ extension can be appreciated by considering a three-phase flow situation depicted in Fig.~\ref{fig_lf_extrap_schematic} (A). Here, a gas bubble is partially trapped in solid metal. The liquid fraction is 1 in the liquid phase and 0 in the solid phase. The liquid fraction value in the gas region is set to zero. For the three-phase surface tension model, the liquid fraction in the diffuse region on the gas side should have the same value as in the surrounding PCM region, as depicted in Fig.~\ref{fig_lf_extrap_schematic} (B). For example, on the bubble-PCM interface, $\varphi$ should equal 1 in the upper half but 0 in the lower half. It is difficult to manually set the liquid fraction within the gas region when interfaces evolve dynamically during simulation. \REVIEW{An approach to address this issue is to solve a hyperbolic equation of the form~\cite{aslam2004partial}}
\begin{equation}
\D{\widetilde{\varphi}}{\tau}+(1-H)\, \n\cdot\grad{\widetilde{\varphi}}=0, 
\label{eq_extrap}
\end{equation}
to achieve this extension. Here, $\widetilde{\varphi}$ is the extended/extrapolated liquid fraction variable and $\n=-\frac{\grad{\REVIEW{\Phi}}}{\left| \grad{\REVIEW{\Phi}} \right|}$ denotes the interfacial normal that points outward from the PCM domain and into the gas phase. This equation extrapolates/extends the liquid fraction $\varphi$ from the PCM to the gas domain along the normal direction $\n$. The factor $(1-H)$ appearing in Eq.~\eqref{eq_extrap} ensures that the extrapolation occurs only one-way, i.e, from PCM to gas. To ensure $\widetilde{\varphi}$ is defined adequately within the smeared region, we extend it up to 5 grid cells within the gas domain. This requires integrating Eq.~\eqref{eq_extrap} for approximately 15 pseudo times steps of size $\Delta \tau = 0.3 \Delta$. Here, $\Delta$ denotes the cell size and a CFL number of 0.3 is employed to ensure convective stability. An example of original and extrapolated liquid fraction variables, $\varphi$ and $\widetilde{\varphi}$, respectively, during a three-phase flow simulation considered in Sec.~\ref{sec_bubble_trap} is shown in Fig.~\ref{fig_lf_extrap_schematic}. We remark that $\widetilde{\varphi}$ is used only to apply surface tension forces along the gas-liquid interface. Wherever else it is applicable, for instance, when defining thermophysical properties (using Eq.~\eqref{eq_EOS}), the original liquid fraction variable $\varphi$ is used.  In light of this discussion, the form of the three-phase surface tension force used in this work reads as
\begin{equation}
\V{f}_{\rm st}^{\rm 3-phase} = \V{f}_{\rm st} \widetilde{\varphi}. \label{eq_st_3phase_lf_extrap}
\end{equation}

\subsection{Heat source}

In most of the problems considered in this work, PCM's phase change is induced by an imposed temperature condition at the domain boundary. There are also cases where phase change is induced by a heat source, which is denoted $Q_\text{src}$ in Eq.~\eqref{eq_enthalpy}. Physically, $Q_\text{src}$ models melting of metals by lasers, electron beams, or electric arcs, which is common in metal additive manufacturing processes~\cite{cook2020simulation}. The dimension of $Q_\text{src}$ is power (Watts) per unit volume. In general, laser power is expressed in Watts (or in terms of heat flux). Through the use of the delta function, an imposed heat flux ($q''$) can be incorporated into the enthalpy equation as
\begin{equation}
Q_{\rm src} = q'' \delta = q''\,|\grad{\widetilde{B}}|.
\label{eq_laser_src}
\end{equation}
As in Eqs.~\eqref{eq_marangoni}, we use a smooth delta function $\delta = |\grad{\widetilde{B}}|$ that is obtained from a mollified Heaviside function $\widetilde{B}$ in Eq.~\eqref{eq_laser_src}.

\section{Discretization}  \label{sec_discretization}

\subsection{Spatial discretization}

\begin{figure}
  \centering
  \subfigure[A 2D staggered Cartesian grid]{
    \includegraphics[scale = 0.45]{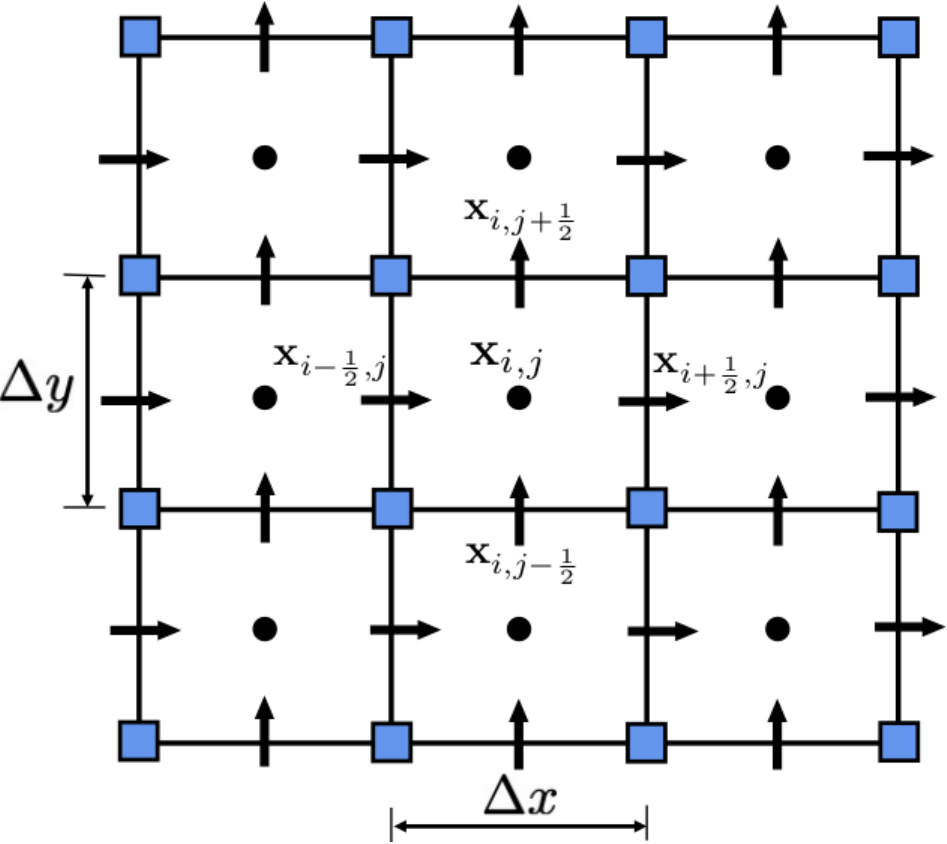} 
    \label{fig_discretized_staggered_grid}
  }
   \subfigure[A single Cartesian grid cell]{
    \includegraphics[scale = 0.35]{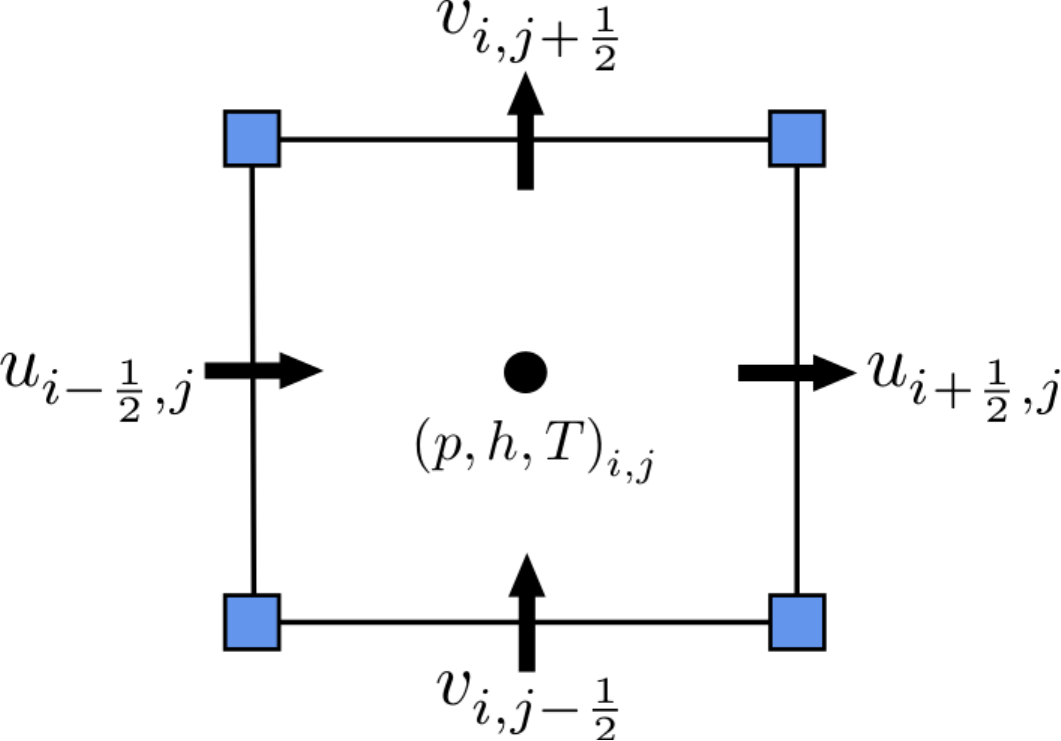}
    \label{fig_single_cell}
  }
  \caption{An illustration of a 2D staggered Cartesian grid. \subref{fig_discretized_staggered_grid} shows the coordinate system for the staggered grid. \subref{fig_single_cell} shows a single grid cell with velocity components $u$ and $v$ approximated at the cell faces (${\bf{\rightarrow}}$) and scalar variables such as pressure $p$, specific enthalpy $h$ and temperature $T$ approximated at the cell center ($\bullet$).}
\label{fig_cfd_grid}
\end{figure}

The continuous equations of motion are discretized on a locally refined staggered Cartesian grid. In this section, we explain the discretization in two spatial dimensions at the coarsest/single level without sacrificing generality. Locally refined grids are addressed in Sec.~\ref{sec_amr}.  

The coarsest level of the grid covers the entire domain $\Omega$ with $\Nx \times \Ny$ rectangular cells. The cell size in the $x$ and $y$ directions is $\dx$ and $\dy$, respectively, as illustrated in Fig.~\ref{fig_discretized_staggered_grid}. Unless stated otherwise, a uniform grid spacing  $\Delta x = \Delta y = \Delta$  is used for all simulations in this work.  We assume that the bottom-left corner of the computational domain $\Omega$ aligns with the origin $(0, 0)$. The position of each center of the grid cell is given by $\x_{i,j} = \left((i + \half)\dx,(j + \half)\dy\right)$, where $i = 0, \ldots, \Nx - 1$ and $j = 0, \ldots, \Ny - 1$. The face center in the $x-$direction, which is located half a grid space away from the cell center $\x_{i,j}$ in the negative $x-$direction, is represented by $\x_{i-\half,j} = \left(i\dx,(j + \half)\dy\right)$, where $i = 0, \ldots, \Nx$ and $j = 0, \ldots, \Ny - 1$. Similar conventions apply to other face center locations. Scalar variables such as pressure ($p$), specific enthalpy ($h$), and temperature ($T$) are stored at the cell centers. The $x-$component of velocity ($u$) is stored at the centers of  $x-$direction cell faces, while the $y-$component of velocity $v$ is stored at the centers of the faces of the $y-$ direction cell, as shown in Fig.~\ref{fig_single_cell}. Material properties, including density ($\rho$), are stored at the cell centers along with interface tracking variables $\REVIEW{\Phi}$ and $\varphi$. Second-order interpolation is used to interpolate the cell-centered quantities to face. Momentum forcing terms, such as surface tension, gravity, and Carman-Kozeny drag force, are stored at the cell faces. Standard second-order finite differences are employed to approximate spatial differential operators. The spatial discretizations of the key continuous operators are as follows:
\begin{itemize}
\item The divergence of the velocity field
$\u = (u,v)$ is approximated at cell centers by
\begin{subequations}
\begin{alignat}{2}
\label{eq_div_fd}
& \vD \cdot \u = D^x u + D^y v, \\
&(D^x u)_{i,j} = \frac{u_{i+\half, j} - u_{i-\half, j}}{\dx}, \\
&(D^y v)_{i,j} = \frac{v_{i, j+\half} - v_{i, j-\half}}{\dy}. 
\end{alignat}
\end{subequations}
\item The diffusion term in the energy equation is approximated as
\begin{align}
\label{eq_energy_diff}
\vD \cdot \left(\kappa \grad{T}\right) = &\frac{1}{\dx}\left(\kappa_{i+\half,j}\frac{T_{i+1,j} - T_{i,j}}{\dx} - \kappa_{i-\half,j}\frac{T_{i,j} - T_{i-1,j}}{\dx}\right) \nonumber \\ 
					     &+\frac{1}{\dy} \left(\kappa_{i,j+\half}\frac{T_{i,j+1} - T_{i,j}}{\dy} - \kappa_{i,j-\half}\frac{T_{i,j} - T_{i,j-1}}{\dy}\right)
\end{align}
\item The gradient of cell-centered quantities (i.e., $p$) is approximated at cell faces by
\begin{subequations}
\begin{alignat}{2}
\label{eq_grad_fd}
& \G p = (G^x p, G^y p), \\
&(G^x p)_{i-\half,j} = \frac{p_{i,j} - p_{i-1,j}}{\dx}, \\
&(G^y v)_{i,j - \half} =\frac{p_{i,j} - p_{i,j-1}}{\dy}. 
\end{alignat}
\end{subequations}
\item \SECONDREVIEW{The continuous strain rate tensor form of the viscous term, which couples the velocity components through spatially variable viscosity, is given by
\begin{equation}
\label{eq_visc_cont}
\div \left[\mu \left(\grad \u + \grad \u^\intercal\right) \right] = \left[
\begin{array}{c}
 (\Lmu \u)^x_{i-\half,j} \\
 (\Lmu \u)^y_{i,j-\half}  \\
\end{array}
\right] = 
\left[
\begin{array}{c}
 2 \D{}{x}\left(\mu \D{u}{x}\right) + \D{}{y}\left(\mu\D{u}{y}+\mu\D{v}{x}\right) \\
 2 \D{}{y}\left(\mu \D{v}{y}\right) + \D{}{x}\left(\mu\D{v}{x}+\mu\D{u}{y}\right) \\
\end{array}
\right].
\end{equation}}
\item The viscous operator is discretized using standard second-order, centered finite differences
\begin{subequations}
\begin{alignat}{2}
 (\Lmu \u)^x_{i-\half,j} &= \frac{2}{\dx}\left[\mu_{i,j}\frac{u_{i+\half,j} - u_{i-\half,j}}{\dx} -
					        \mu_{i-1,j}\frac{u_{i-\half,j} - u_{i-\3half,j}}{\dx}\right] \nonumber \\ 
                    &+ \frac{1}{\dy}\left[\mu_{i-\half, j+\half}\frac{u_{i-\half,j+1} - u_{i-\half,j}}{\dy} - 
					         \mu_{i-\half, j-\half}\frac{u_{i-\half,j} - u_{i-\half,j-1}}{\dy}\right] \nonumber \\
	            &+ \frac{1}{\dy}\left[\mu_{i-\half, j+\half}\frac{v_{i,j+\half} - v_{i-1,j+\half}}{\dx} - 
					         \mu_{i-\half, j-\half}\frac{v_{i,j-\half} - v_{i-1,j-\half}}{\dx}\right] \label{eq_viscx_fd} \\				         
 (\Lmu \u)^y_{i,j-\half} &= \frac{2}{\dy}\left[\mu_{i,j}\frac{v_{i,j+\half} - v_{i,j-\half}}{\dy} -
					        \mu_{i,j-1}\frac{v_{i,j-\half} - v_{i,j-\3half}}{\dy}\right] \nonumber \\ 
                    &+ \frac{1}{\dx}\left[\mu_{i+\half, j-\half}\frac{v_{i+1,j-\half} - v_{i,j-\half}}{\dx} - 
					         \mu_{i-\half, j-\half}\frac{v_{i,j-\half} - v_{i-1,j-\half}}{\dx}\right] \nonumber \\
	            &+ \frac{1}{\dx}\left[\mu_{i+\half, j-\half}\frac{u_{i+\half,j} - u_{i+\half,j-1}}{\dy} - 
					         \mu_{i-\half, j-\half}\frac{u_{i-\half,j} - u_{i-\half,j-1}}{\dy}\right] \label{eq_viscy_fd},
\end{alignat}
\end{subequations}
in which viscosity is required at both cell centers and nodes of the staggered grid (i.e., $ \mu_{i\pm\half, j\pm\half}$).
Node-centered quantities are obtained via interpolation by either arithmetically or harmonically averaging the neighboring cell-centered quantities. In this work, we utilize harmonic averaging.
\end{itemize}

 In terms of convective discretization, we employ the third-order accurate cubic upwind interpolation (CUI) scheme. The CUI method satisfies both the convection-boundedness criterion (CBC) and the total variation diminishing (TVD) property. Specifically, the CUI scheme demonstrates third-order spatial accuracy in monotonic regions, where the gradient of the advected quantity remains monotone, and transitions to first-order spatial accuracy in non-monotonic regions due to upwinding. We do not present the spatial discretization of the advection equation using CUI here for brevity, but detailed information can be found in our prior publication \cite{nangia2019robust}. Extending these discretizations to three-dimensional Cartesian grids  is straightforward. For further details on the spatial discretization and boundary conditions on adaptively refined meshes, refer our prior works~\cite{nangia2019robust, Nangia2019WSI, Bhalla13}.

 \subsection{Temporal discretization} \label{sec_temp_discretization}
In this section, we detail the temporal discretization method utilized in our study for the continuous equations of motion. Within each time step [$t^n, t^{n+1}$], we employ $\ncycles$ cycles of fixed-point iteration to approximate the solution to the coupled fluid-thermal system. At the start of the simulation ($t=0$), all variables are initialized with the initial conditions of the problem. At the beginning of the each time step, with cycle number $k=0$, variables are initialized to their values from the previous time step, denoted as $\u^{n+1,0} = \u^{n}$, $p^{n+\half,0} = p^{n-\half}$, $h^{n+1,0} = h^{n}$, $\REVIEW{\Phi}^{n+1,0} = \REVIEW{\Phi}^{n}$, and $\varphi^{n+1,0} = \varphi^{n}$, and iterated till $k=\ncycles-1$.  Convective and surface tensions terms are treated explicitly, while other terms are treated implicitly. Unless otherwise specified, $\ncycles = 2$ is used in this work.

 
\subsubsection{Consistent mass-momentum integrators} \label{sec_cons_mass_mom}

For incompressible multiphase flows, mass is advected indirectly via an interface tracking variable. In the present work this corresponds to the advection of the level set variable $\REVIEW{\Phi}$. The momentum equation also contains a mass flux term $\V{m_\rho} \equiv \rho \u$ in the the convective operator $\div (\rho \u \otimes \u) = \div(\V{m_\rho} \otimes \u)$. For high density ratio flows $\rhol/\rhog > 100$ this weak coupling of mass and momentum leads to numerical instabilities. One way to ensure strong coupling between mass and momentum advection is to solve the redundant mass balance Eq.~\eqref{eq_mass_blance} and use the same mass flux $\V{m_\rho}$ in the two advective operators: $\div(\rho \u)$ and $\div (\rho \u \otimes \u)$. \REVIEW{This idea was proposed by Desjardins and Moureau~\cite{desjardins2010methods} in 2010 and later extended to achieve second-order accuracy in our prior works [29,30]}. 

Although the use of the same discrete mass flux is essential for ensuring stability of high density ratio flows, there is still some freedom in the choice of time integrators for mass and momentum equations. In our prior works we employed  the strong stability preserving Runga-Kutta (SSP-RK3) scheme for solving the mass balance equation and a midpoint RK-2 scheme for the momentum equation. Specifically the mass balance equation is integrated in three stages:
\begin{subequations} \label{eq_ssp}
 \begin{alignat}{2} 
& \breve{\vrho}^{(1)} = \breve{\vrho}^{n} - \dt \vec{M} \left(\u_{\rm adv}^n,  \breve{\vrho}_{\rm lim}^{n}\right), \label{eq_ssp_first_stage} \\
& \breve{\vrho}^{(2)} = \frac{3}{4}\breve{\vrho}^{n} + \frac{1}{4}\breve{\vrho}^{1} -\frac{1}{4} \dt \vec{M} \left(\u_{\rm adv}^{(1)},  \breve{\vrho}_{\rm lim}^{(1)}\right), \label{eq_ssp_second_stage} \\
& \breve{\vrho}^{n+1, k+1} = \frac{1}{3}\breve{\vrho}^{n} + \frac{2}{3}\breve{\vrho}^{(2)} -\frac{2}{3} \dt \vec{M} \left(\u_{\rm adv}^{(2)},  \breve{\vrho}_{\rm lim}^{(2)}\right), \label{eq_ssp_last_stage}
\end{alignat} 
\end{subequations}
in which $ \breve{\vrho}$ denotes the face-centered density and $\u_{\rm adv}$ represents the advection velocity that is centered at the faces of the staggered (velocity) control volumes. The right-hand side term in Eqs.~\eqref{eq_ssp} $ \vec{M} \left(\u_{\rm adv},  \breve{\vrho}_{\rm lim}\right) \approx \left[ \left(\div(\u_{\rm adv}  \breve{\vrho}_{\rm lim}) \right)_{i-\frac{1}{2}, j},   \left(\div(\u_{\rm adv}  \breve{\vrho}_{\rm lim}) \right)_{i, j-\frac{1}{2}}\right]$ is an explicit CUI limited approximation to the linear advection term $\div(\rho \u)$. Subscripts ``lim" and ``adv" indicate a limited and an advected quantity, respectively. To maintain the accuracy of the SSP-RK3 integrator, it is essential to use  velocities that are appropriately extrapolated or interpolated in time. To wit, for the first cycle ($k=0$), the velocities are given by
\begin{subequations} \label{eq_vsspk0}
 \begin{alignat}{2}
& \u^{(1)} = 2\u^{n} -  \u^{n-1}, \\
& \u^{(2)} = \frac{3}{2}\u^{n} - \frac{1}{2}\u^{n-1},
\end{alignat} 
\end{subequations}
and for the remaining cycles ($k>0$), the velocities are
\begin{subequations} \label{eq_vsspk+}
 \begin{alignat}{2}
& \u^{(1)} = \u^{n+1,k}, \\
& \u^{(2)} = \frac{3}{8}\u^{n+1,k} + \frac{3}{4}\u^{n}-\frac{1}{8}\u^{n-1}.
\end{alignat} 
\end{subequations}

The consistency between mass and momentum advection is achieved by using the same discrete mass flux $ \V{m_\rho} =   \breve{\vrho}_{\rm lim}^{(2)} \u_{\rm adv}^{(2)} $ from the last stage of the SSP-RK3 integrator (Eq.~$\ref{eq_ssp_last_stage}$) in the discrete momentum equation
 \begin{align}
 \frac{\breve{\vrho}^{n+1, k+1}  \u^{n+1, k+1} - \breve{\vrho}^{n}  \u^{n}}{\Delta t}+\mathbf{C} \left(\u_{\rm adv}^{(2)}, \breve{\vrho}_{\rm lim}^{(2)} \u_{\rm lim}^{(2)} \right) &= -\G p^{n+\frac{1}{2}, k+1} + (\L_\mu \u)^{n+\frac{1}{2}, k+1} \nonumber \\ 
 &+\breve{\vrho}^{n+1, k+1} \g - (A_d \u)^{n+1, k+1} +  \f_{\rm st}^{{\rm 3-phase}, n+\frac{1}{2}, k+1}.
\label{eq_adv_ns}
\end{align} 
in which the convective term is approximated as
 \begin{align}
\mathbf{C} \left(\u_{\rm adv}^{(2)}, \breve{\vrho}_{\rm lim}^{(2)} \u_{\rm lim}^{(2)} \right)& \approx \left[ \left(\div{\left(\u_{\rm adv}^{(2)} \breve{\vrho}_{\rm lim}^{(2)} u_{\rm lim}^{(2)}\right)} \right)_{i-\frac{1}{2}, j},  \left(\div{\left(\u_{\rm adv}^{(2)} \breve{\vrho}_{\rm lim}^{(2)} v_{\rm lim}^{(2)}\right)} \right)_{i, j- \frac{1}{2}}\right],
\label{eq_convective_term}
\end{align} 
and $\breve{\vrho}^{n+1, k+1}$ is the updated value of the density obtained from the last stage of the SSP-RK3 integrator. The term $A_d = C_d \frac{\displaystyle \varphi_\text{S}^2}{\displaystyle (1- \varphi_\text{S})^3 + \epsilon}$ in the momentum equation represents the Carman-Kozeny drag coefficient that is used to retard any flow in the solid domain. Here, $\varphi_\text{S} = H(1 -  \varphi)$ is the solid fraction of the grid cell, and $\epsilon = 10^{-3}$ is a small number to avoid
a division by zero and to control the strength of penalty parameter
($C_d/\epsilon$) in the solid region. In the solid domain, $C_d$ is also taken to be large in magnitude to retard fluid motion. When equating drag force with the first term on the left hand side of the momentum equation, a sufficiently large value is obtained for $C_d = \rhos/\Delta t$.  The three-phase surface tension force is evaluated using the midpoint values of the level set function $\REVIEW{\Phi}^{n+\half, k+1} = \half(\REVIEW{\Phi}^{n+1, k+1} + \REVIEW{\Phi}^{n})$. 

In our prior works, for the test problems considered, it was  demonstrated that although the mass and momentum equations employed different time integrators, the use of the same discrete mass flux was sufficient to ensure numerical stability. In this work, we propose an additional stabilizing term in the momentum equation to account for different integration schemes used for mass and momentum equations. Specifically, after integrating the mass balance equation (using SSP-RK3), the (face-centered) residual of the equation is computed 
\begin{align}
 \V{\mathcal{R}} = \frac{\breve{\vrho}^{n+1, k+1}   - \breve{\vrho}^{n}}{\Delta t}+\vec{M} \left(\u_{\rm adv}^{(2)},  \breve{\vrho}_{\rm lim}^{(2)}\right). 
\label{eq_mass_residual}
\end{align}  
The  residual $ \V{\mathcal{R}}$ is expected to be very small (close to zero). The residual of the mass equation is used to define a body force in the momentum equation. The modified momentum equation reads as
 \begin{align}
 \frac{\breve{\vrho}^{n+1, k+1}  \u^{n+1, k+1} - \breve{\vrho}^{n}  \u^{n}}{\Delta t}+\mathbf{C} \left(\u_{\rm adv}^{(2)}, \breve{\vrho}_{\rm lim}^{(2)} \u_{\rm lim}^{(2)} \right) &=   \V{\mathcal{R}} \u^{n+1,k} + \text{ viscous + pressure + other forces}.
\label{eq_mom_residual}
\end{align} 

The rationale for adding $\V{\mathcal{R}}  \u^{n+1,k}$ to the right hand side of the momentum equation is as follows. Consider a uniform and constant velocity ($\u \equiv \u_c$) flow which is not subject to any viscous, pressure or body force. In such a case the momentum equation reverts to the mass balance equation:
\begin{subequations}
\begin{alignat}{2}
&\D{\left(\rho \u\right)}{t}+\div{\left(\rho \u \otimes \u\right)} = 0 \\
\hookrightarrow & \u_c \left( \D{\rho}{t} + \div (\rho \u_c) \right) = 0, \\
\hookrightarrow  & \u_c \left( \D{\rho}{t} + \div (\rho \u)  \right) = \V{\mathcal{R}}\u_c  = 0. 
\end{alignat}
\end{subequations}
Thus the addition of the close-to-zero term $\V{\mathcal{R}}  \u^{n+1,k}$ ensures that the left and right hand sides of the modified momentum Eq.~\eqref{eq_mom_residual} discretely balance each other. 

Another way to ensure consistency between mass and momentum transport is to employ the same integration scheme for the two equations. Since our existing fluid solvers employ a midpoint/RK-2 scheme for integrating the momentum equation, we choose to employ the same integration scheme for the mass equation. In this case, the density update reads as:
 \begin{align}
\label{eq_rk2}
& \breve{\vrho}^{n+1, k+1} = \vrho^{n} - \dt \vec{M} \left(\u_{\rm adv}^{(1)},  \breve{\vrho}_{\rm lim}^{(1)}\right),
\end{align} 
in which $\u_{\rm adv}^{(1)} = \frac{1}{2} \left(\u_{\rm adv}^{n} + \u_{\rm adv}^{n+1, k} \right)$ and $ \breve{\vrho}_{\rm lim}^{(1)} =  \frac{1}{2} \left( \vrho_{\rm lim}^{n} + \breve{\vrho}_{\rm lim}^{n+1, k}\right)$. It can be noted that the above scheme reduces to forward Euler (RK1) scheme with $\ncycles = 1$. With the same integration scheme for mass and momentum equations (RK-2) it is not necessary to add $\V{\mathcal{R}}  \u^{n+1,k}$ to the right hand side of the momentum equation. In our original low Mach enthalpy work~\cite{thirumalaisamy2023lowmach} we employed the RK-2 integrator for both mass and momentum equations without considering the residual term. In this work and in our revised code, we now retain the extra stabilizing term in the momentum equation. 


\vspace{1em}
\noindent \textbf{\underline{A test problem to check consistency of mass-momentum transport:}} To illustrate the importance of consistent mass and momentum transport for high density ratio flows, consider an advection of a dense, isothermal \REVIEW{droplet}\footnote{\REVIEW{This benchmark test is provided in IBAMR GitHub within the directory \texttt{examples/phase\_change/ex2}.}} moving in a uniform velocity $\u_c = (u, v) = (1, 1)$ in a square periodic domain $\Omega \in [0, 1]^2$. The radius of the \REVIEW{droplet} is $R=0.2$ and it is initially centered at the middle of the domain: $(x_i=0.5, y_i=0.5)$. The density ratio between the \REVIEW{droplet} and the outer liquid is $\rho_i/\rho_o = 10,000$. The domain is discretized with 2 levels of mesh, with the coarsest mesh size of $N_x \times N_y = 128^2$. A refinement ratio of $n_{\rm ref} = 2$ is used between the two levels. The \REVIEW{droplet}'s interface is always placed on the finest level. Further details on the adaptive mesh refinement (AMR) are provided in the next section. Viscosity $\mu$ is set to zero in the entire domain, and no other body force (e.g., gravity or surface tension) is present. Under these conditions, the governing equations for this problem reduce to
\begin{subequations}
\begin{alignat}{2}
&\D{\REVIEW{\Phi}}{t}+\u\cdot\grad{\REVIEW{\Phi}}=0, \\
& \D{\rho}{t}+\div{\left(\rho \u\right)}=0, \\
&\D{\left(\rho \u\right)}{t}+\div{\left(\rho \u \otimes \u\right)} = 0.
\end{alignat}
\end{subequations}
For this problem it is expected that the \REVIEW{droplet} advects with a constant velocity $\u_c$ and returns to its original position (with minimal distortions) after $\tau = d/|\u_c|$, in which $d$ represents the distance travelled to reach its original position. We simulate the advection of the \REVIEW{droplet} considering four scenarios:
\begin{itemize}
\item Case~A: RK-2 integrator for momentum equation and SSP-RK3 integrator for mass balance equation.
\item Case~B: RK-2 integrator for momentum equation, SSP-RK3 integrator for mass balance equation, and  residual force $\mathcal{R} \u$ in the momentum equation.
\item Case~C: RK-2 integrator for both mass and momentum equations.
\item Case~D: RK-2 integrator for both mass and momentum equations, and residual force $\mathcal{R} \u$ in the momentum equation.
\end{itemize}
In all four cases we consider two cycles of fixed-point iteration, i.e., $\ncycles = 2$ is used. To assess the performance of the integrator choice, we compute the percentage change in mass and momentum of the system. The relative change is defined as
\begin{equation}
\mathcal{E}_\mathcal{B} = \frac{| \beta(t) - \beta_0|}{\beta_0}\times 100,
\label{eq_rel_change}
\end{equation}
in which $\beta =  \int_\Omega \rho\, \rm{d}V$ and $\mathcal{B} = \mathcal{M}$ denote the relative change in the mass of the system,  and $\beta =  \int_\Omega \rho \u\, \rm{d} V$ and $\mathcal{B} = \mathcal{L}_{\x}$ denote the relative change in the linear momentum of the system. \REVIEW{$\beta_0$ represents quantities computed using the system's initial conditions.}  For this problem we expect that the mass and momentum of the system should remain invariant over time. In order to check whether any catastrophic instability occurs in the four cases, we run the simulations till $t = 10$, which corresponds to 10 complete passes of the \REVIEW{droplet} across the domain.   

\begin{figure}
\centering
\includegraphics[width=0.95\linewidth]{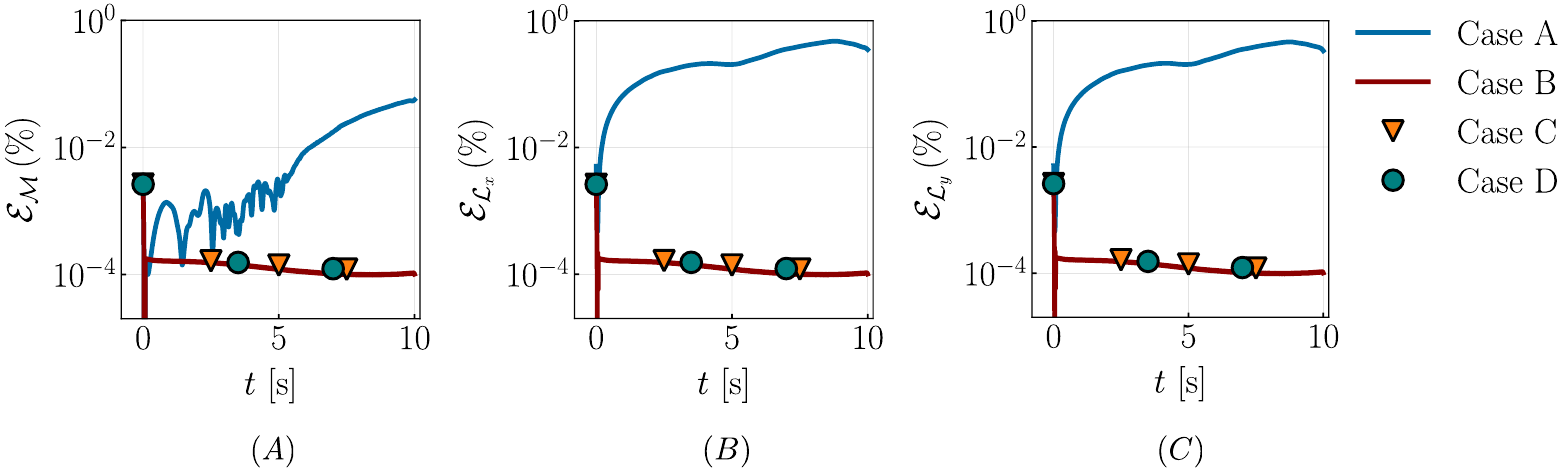}
\caption{Percentage change in (A) mass (B) x-momentum and (C) the y-momentum of the system for various choices of time integrators: Case~A - RK-2 integrator for momentum equation and SSP-RK3 integrator for mass balance equation;  
Case~B - RK-2 integrator for momentum equation, SSP-RK3 integrator for mass balance equation, and  residual force $\mathcal{R} \u$ in the momentum equation; Case~C - RK-2 integrator for both mass and momentum equations; Case~D - RK-2 integrator for both mass and momentum equations, and residual force $\mathcal{R} \u$ in the momentum equation. $\ncycles = 2$ is used for all four cases. }
\label{fig_bubble_adv_residual_effect}
\end{figure}

\begin{figure}
\centering
\includegraphics[width=0.95\linewidth]{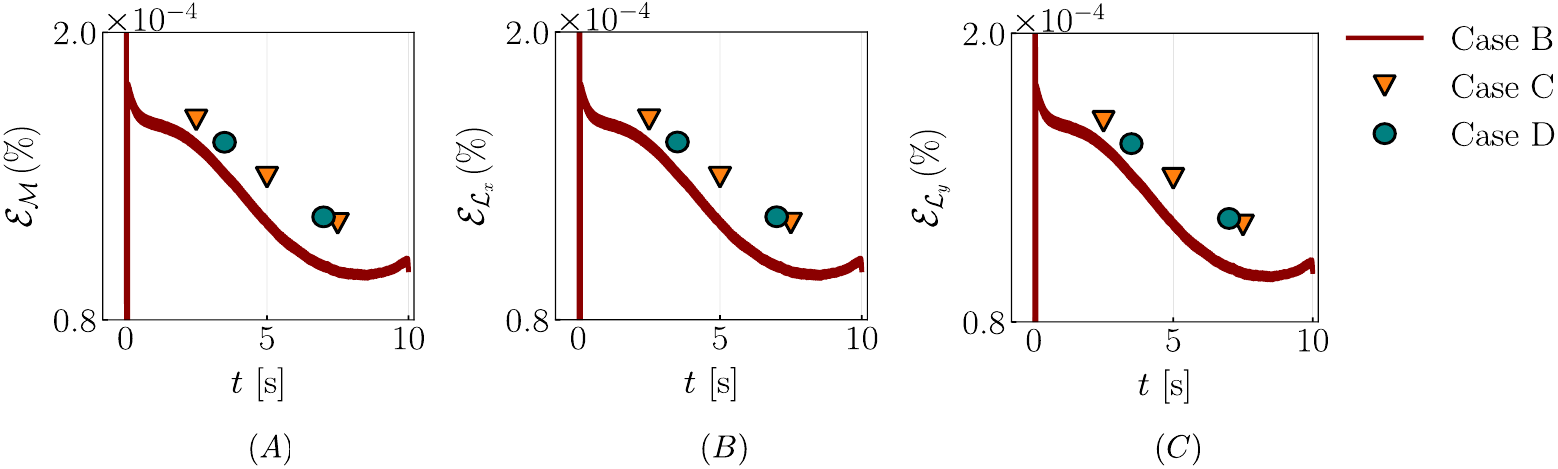}
\caption{\REVIEW{Percentage change in (A) mass (B) x-momentum and (C) the y-momentum of the system for various choices of time integrators:   
Case~B - RK-2 integrator for momentum equation, SSP-RK3 integrator for mass balance equation, and  residual force $\mathcal{R} \u$ in the momentum equation; Case~C - RK-2 integrator for both mass and momentum equations; Case~D - RK-2 integrator for both mass and momentum equations, and residual force $\mathcal{R} \u$ in the momentum equation. $\ncycles = 2$ is used for all four cases.} }
\label{fig_bubble_adv_residual_effect_wo_caseA}
\end{figure}

Fig.~\ref{fig_bubble_adv_residual_effect} shows the results for the four cases. It can be observed that Case~A produces spurious changes in mass and momentum of the system that increase slowly over time. However, adding the residual force in the momentum equation (Case~B) reduces the spurious changes in mass and momentum, and keeps them relatively low and constant over time. Similarly, using the same time integrator (Case~C and Case~D) also yields very small percentage changes in both mass and momentum of the system that remain steady over time. \REVIEW{The spurious changes in mass and momentum for Cases B, C, and D have similar magnitudes, as shown in Fig.~\ref{fig_bubble_adv_residual_effect_wo_caseA}.} In all four cases, the simulation remained stable till $t = 10$. The dynamics of the problem are illustrated in Fig.~\ref{fig_bubble_adv_dynamics} of the results Sec.~\ref{sec_results}. Based on the results of this test, we choose the integrator combination of Case~D (i.e., RK-2 integrator for integrating both mass and momentum equations and considering an additional residual force in the momentum equation) in this work. 

\subsubsection{Consistent mass-momentum-enthalpy integrators}

The concept of consistent mass-momentum transport is extended to consistent mass-momentum-enthalpy transport in this section. This involves solving the mass balance equation on both cell faces and cell centers. The former is done to obtain the discrete mass flux $\V{m_\rho}^\text{fc}$ for the convective operator $\div (\V{m_\rho}^\text{fc} \otimes \u)$ of the momentum equation, while the latter is done to obtain the discrete mass flux $\V{m_\rho}^\text{cc}$ for the advective operator $\div (\V{m_\rho}^\text{cc} h)$ of the enthalpy equation. We use an RK-2 scheme for time integrating both cell-centered mass and enthalpy equations, as explained in the previous section. An additional residual term $\mathcal{R} h^{n+1,k}$ is added to the right hand side of the enthalpy equation for reasons explained previously. Overall the discretized cell-centered enthalpy equation reads as
\begin{align}
 \frac{\breve{\vrho}^{n+1, k+1}  h^{n+1, k+1} - \breve{\vrho}^{n}  h^{n}}{\Delta t}+ A \left(h_{\rm lim}^{(1)}, \breve{\vrho}_{\rm lim}^{(1)} \u_{\rm adv}^{(1)} \right)   &= \mathcal{R} h^{n+1,k} + (\div\kappa\grad T)^{n+1,k+1} +  Q_{\rm src}^{n+1,k},
\label{eq_enthalpy_discrete}
\end{align} 
in which the advective term is defined as 
\begin{align}
A \left(h_{\rm lim}^{(1)}, \breve{\vrho}_{\rm lim}^{(1)} \u_{\rm adv}^{(1)} \right)& \approx \left[ \left(\div{\left(h_{\rm lim}^{(1)} \breve{\vrho}_{\rm lim}^{(1)} u_{\rm adv}^{(1)}\right)} \right)_{i-\frac{1}{2}, j},  \left(\div{\left(h_{\rm lim}^{(1)} \breve{\vrho}_{\rm lim}^{(1)} v_{\rm adv}^{(1)}\right)} \right)_{i, j- \frac{1}{2}}\right].
\label{eq_h_convective_term}
\end{align} 
In Eq.~\eqref{eq_enthalpy_discrete} $\mathcal{R}$ is the (cell-centered) residual of the discrete mass balance equation (see also Eq.~\eqref{eq_mass_residual}). We explore the combined consistency of mass, momentum, and enthalpy integrators in Sec.~\ref{sec_bubble advection_enthalpy}. 


\subsubsection{Solving the non-linear enthalpy equation} \label{sec_enthalpy_newton}
The enthalpy equation is a non-linear equation since thermophysical properties like $\kappa$ depend on liquid fraction $\varphi$, which in turn is a function of enthalpy $h$. The non-linear equation is solved using Newton's method. A Newton iteration is denoted by a superscript $m$, where $m = 0,\ldots,q_\text{max}-1$. Within the $k^\text{th}$-cycle, Newton iterations begin by initializing $h^{n+1,k,0} = h^{n+1,k}$, $T^{n+1,k,0} = T^{n+1,k}$, and $\varphi^{n+1,k,0} = \varphi^{n+1,k}$. The iterations proceed until a suitable convergence criterion (defined below) is met or the maximum number of iterations $q_\text{max}$ is reached. Newton iterations involve the following steps: 
\begin{enumerate}
    \item Linearize $h^{n+1,k+1}$  using Taylor series expansion
    \begin{equation}
    h^{n+1,k+1,m+1} = h^{n+1,k+1,m} + \D{h}{T}\bigg|^{n+1,k+1,m}(T^{n+1,k+1,m+1}-T^{n+1,k+1,m}),
    \label{eq_enthalpy_linearization}
    \end{equation}
    in which the derivative $\D{h}{T}$ is given by
    \begin{align}
    \D{h}{T}\Bigg|_\text{PCM} = \begin{cases}
     \cps,&  T<\Tsol,\\ 
     \bar{C}+L/(\Tliq-\Tsol),&\Tsol \le T \le \Tliq, \\ 
     \cpl,& T> \Tliq ,
    \end{cases}
    \label{eq_dhdt_pcm}
    \end{align}
    in the PCM domain and $\D{h}{T}\Big|_\text{gas} = \cpg$ in the gas domain.  
    

    \item Substitute $h^{n+1,k+1,m+1}$ from Eq.~\eqref{eq_enthalpy_linearization} into Eq.~\eqref{eq_enthalpy_discrete} and solve for temperature $T^{n+1,k+1,m+1}$.

    \item Update $h^{n+1,k+1,m+1}$ using the Taylor series (Eq.~\eqref{eq_enthalpy_linearization}) and $T^{n+1,k+1,m+1}$.
    
    \item Synchronize $T^{n+1,k+1,m+1}$ and $\varphi^{n+1,k+1,m+1}$ with the updated enthalpy $h^{n+1,k+1,m+1}$. Analytical relations $T$-$h$ and $\varphi$-$h$ written in Eqs.~\eqref{eq_T_pcm}-\eqref{eq_liquid_fraction} are used to synchronize temperature and liquid fraction with enthalpy.

    \item Update thermophysical properties ($\kappa$, $\mu$,$\rho$) using $\varphi^{n+1,k+1,m+1}$ and $H^{n+1, k+1}$ using the mixture model Eq.~\eqref{eq_material_properties}.

    \item Compute the relative change in liquid fraction $\varphi$     
    \begin{equation}
    \Delta_\varphi = \frac{||\varphi^{n+1,k+1,m+1}- \varphi^{n+1,k+1,m}||_2}{1 + || \varphi^{n+1,k+1,m}||_2}.
    \end{equation}
    The Newton solver is deemed to be converged if $\Delta_\varphi  \le 10^{-8}$ or if $m+1 = q_\text{max} = 5$ iterations have completed.  
    
\end{enumerate}

\section{Adaptive mesh refinement (AMR)} \label{sec_amr}
We use a structured adaptive mesh refinement framework to discretize the governing equations. The two-dimensional computational domain $\Omega \in [0, L]\times [0, H]$ is discretized into multiple levels of structured grids. The ratio between each successive grid level is denoted $n_{\rm ref}$. The cell dimensions on the coarsest level are $\Delta x_0 = \frac{L}{N_{x0}}$ and $\Delta y_0 = \frac{H}{N_{y0}}$, in which $N_{x0}$ and $N_{y0}$ are the number of cells in the $x$ and $y$ direction, respectively. The cell dimensions on finer levels are then $\Delta \x_{\rm min} = \Delta \x_{0}/n_{\rm ref}^{l-1}$, in which $l$ is the grid level number. 

We support both static and adaptive meshes in our framework. During static refinement, coarse mesh levels are refined over a fixed area of space, and mesh configuration remains constant over time. It is applicable when the region of interest exhibits minimal movement. With dynamic meshes (or adaptive meshes), coarse level cells are tagged/untagged throughout the simulation based on user-specified criteria. In our framework, we employ two tagging criteria:
\begin{enumerate}

\item Tagging cells based on the signed distance function: To resolve the gas-PCM interface with sufficient mesh resolution, a \emph{value-based tagging} criterion is used. A cell is tagged for grid refinement when its signed distance function value is within zero threshold, e.g., $|\REVIEW{\Phi}| \le 2\Delta x_0$. Using numerical simulations, we find that this amount of tagging is sufficient to resolve the sharp variation in material properties ($\rho$, $\kappa$, $\mu$) across the gas-PCM interface.   

\item Tagging cells based on the liquid fraction variable: In order to capture the mushy region and the associated volume change of PCM, it is necessary to use sufficiently refined grids to resolve the $\Omegam(t)$ region. We can identify the mushy region by either: (i) probing the value of $\varphi$, e.g.,   $0.3 \le \varphi \le 0.8$; or by (ii) probing the gradient of $\varphi$, e.g., $|\grad \varphi| > 0$. We term the former tagging criterion as  \emph{$\varphi$-based tagging} or \emph{value-based tagging} and the latter as \emph{$\grad \varphi$-based tagging} or \emph{gradient-based tagging.}   
\end{enumerate}

\vspace{1em}
\noindent \textbf{\underline{Testing the robustness of AMR in capturing mushy regions:}} Prior works~\cite{nangia2019robust,bhalla2020entryexit, khedkar2021iswec} have confirmed our AMR framework's ability to adaptively refine/coarsen the grid near gas-liquid/gas-solid interfaces in isothermal flows without any phase change. Here we test its capability for the more complex scenario: capturing the evolving mushy region during phase change. We revisit the Stefan problem with expansion from our previous work that introduced the low Mach enthalpy method, but this time using adaptively refined AMR grids.

 The Stefan problem is simulated by considering a quasi one-dimensional computational domain $\Omega \in [0,1] \times [0, 0.05]$ with $l = 2$ mesh levels and a refinement ratio of $n_\text{ref} = 2$\footnote{\REVIEW{This benchmark test is provided in IBAMR GitHub within the directory \texttt{examples/phase\_change/ex1}.}}. The coarsest level ($l = 1$) is covered by $N_{x0} \times N_{y0} = 640 \times 32$ number of cells. The simulation runs until $t=10$ s  with a constant time step size of $\Delta t = 10^{-4}$ s. The domain is periodic in the $y$-direction. Initially, the liquid PCM occupies the entire domain and has a temperature of $T_i = 973.6$ K. The temperature at the left boundary ($x=0$) is set at $T_o = 298.6$ K to induce solidification. The melting/fusion temperature of the PCM is $T_m = 933.6$ K.  The right end of the boundary ($x=1$) is kept adiabatic (homogeneous Neumann boundary condition for temperature). The flow solver uses zero-velocity and zero-pressure/outflow boundary conditions at the left and right ends, respectively. The phase change material's thermophysical properties, largely aluminum-based, are listed in Table~\ref{tab_aluminum_properties}. Here, the liquid and solid densities are assumed to be $\rhol = 2700$ and $\rhos = 500$  kg/m$^3$, respectively, and both fluid and solid viscosities are set to zero. Upon solidification, the material expands due to its lower density than the liquid phase. This problem has been solved both analytically and numerically (on uniform grids) in our previous work~\cite{thirumalaisamy2023lowmach}. 
 We use this problem to analyze the performance of  the two tagging methods ($\varphi$- vs. $\grad \varphi$-based) to identify the mushy region effectively within the AMR framework.

\begin{table}[]
\centering
\caption{Thermophysical properties used to simulate the Stefan problem}
\label{tab_aluminum_properties}
\begin{tabular}{ll}
Property & Value \\
\midrule
Thermal conductivity of solid   $\ks$     &   211  W/m.K  \\
Thermal conductivity of liquid  $\kl$                 & 91  W/m.K    \\
Specific heat of solid  $\cps$                 & 910 J/kg.K      \\
Specific heat of liquid  $\cpl$                  &  1042.4  J/kg.K   \\
Solidification temperature  $T_m$                &  933.6 K     \\
Bulk phase change temperature $T_r$                   & 933.6 K \\
Liquidus temperature $\Tliq$                       & 938.6 K \\
Solidus temperature $\Tsol$                       & 928.6 K  \\
Latent heat  $L$                &  383840  J/kg    \\
\bottomrule
 \end{tabular}
\end{table}

\begin{figure}
\centering
\includegraphics[width=0.9\linewidth]{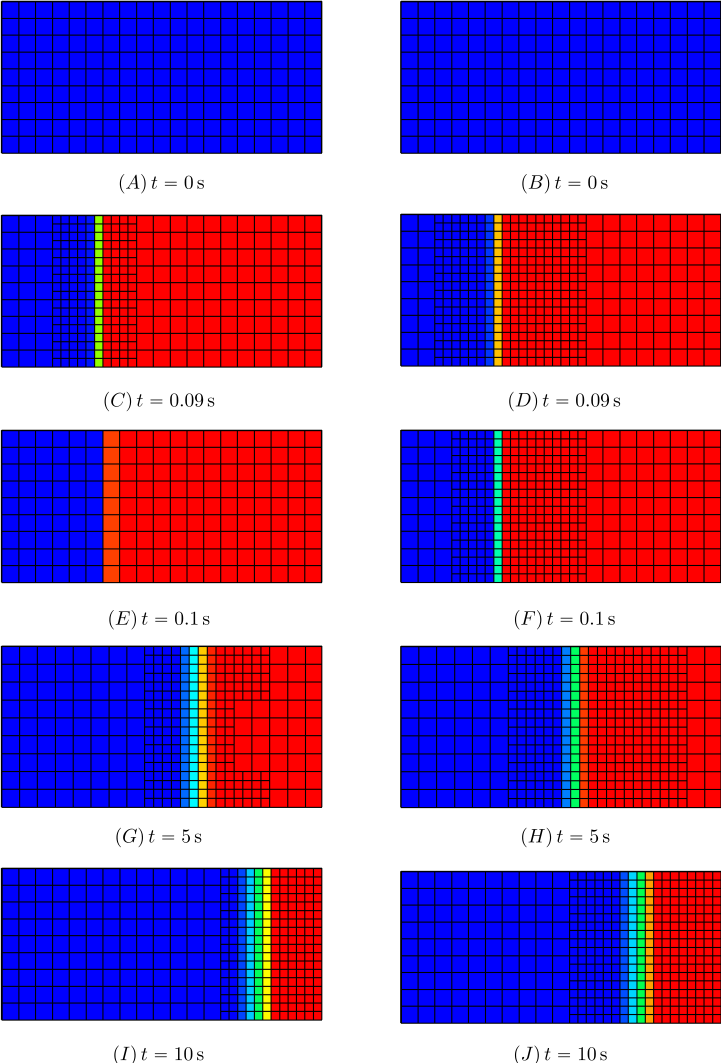}
\caption{Stefan problem with solidification: time  evolution of the liquid fraction $\varphi$ and adaptively refined grids. A low temperature at the left boundary $T_0 < T_m$ induces solidification of the PCM. AMR grid pattern for $\varphi$-based (left) and $\grad{\varphi}$-based (right) tagging approaches. The solid and liquid regions are depicted by blue and red colors, respectively. The narrow region between the all solid and liquid is the mushy zone. }
\label{fig_stefan_problem_expansion}
\end{figure}

The position of the moving (from left to right) solidification front is denoted $x^* = s(t)$. The numerical solid–liquid interface fraction is defined by the iso-contour value 0.5 of the liquid fraction $\varphi$. Fig.~\ref{fig_stefan_problem_expansion} shows the evolution of $\varphi$ over time $t$ using $\varphi$-based (left) and $\grad{\varphi}$-based (right) tagging. It can be seen that the former tagging approach produces intermittent fine mesh patterns. This can be explained as follows. There are instances during the simulation in which the mushy region narrows and falls into the subgrid level. As a result, no cells are tagged using the value-based tagging method. This results in the simulation removing the entire fine grid level $l = 2$. Fig.~\ref{fig_stefan_problem_expansion} (E) illustrates this effect, with $\varphi$ either 0 or 1 and not meeting the value-based refinement criterion. Gradient-based tagging, on the other hand, produces finer meshes near the liquid-solid interface. Despite the narrowness of the mushy region, the gradient of liquid fraction always exists.

\begin{figure}
\centering
\includegraphics[width=0.7\linewidth]{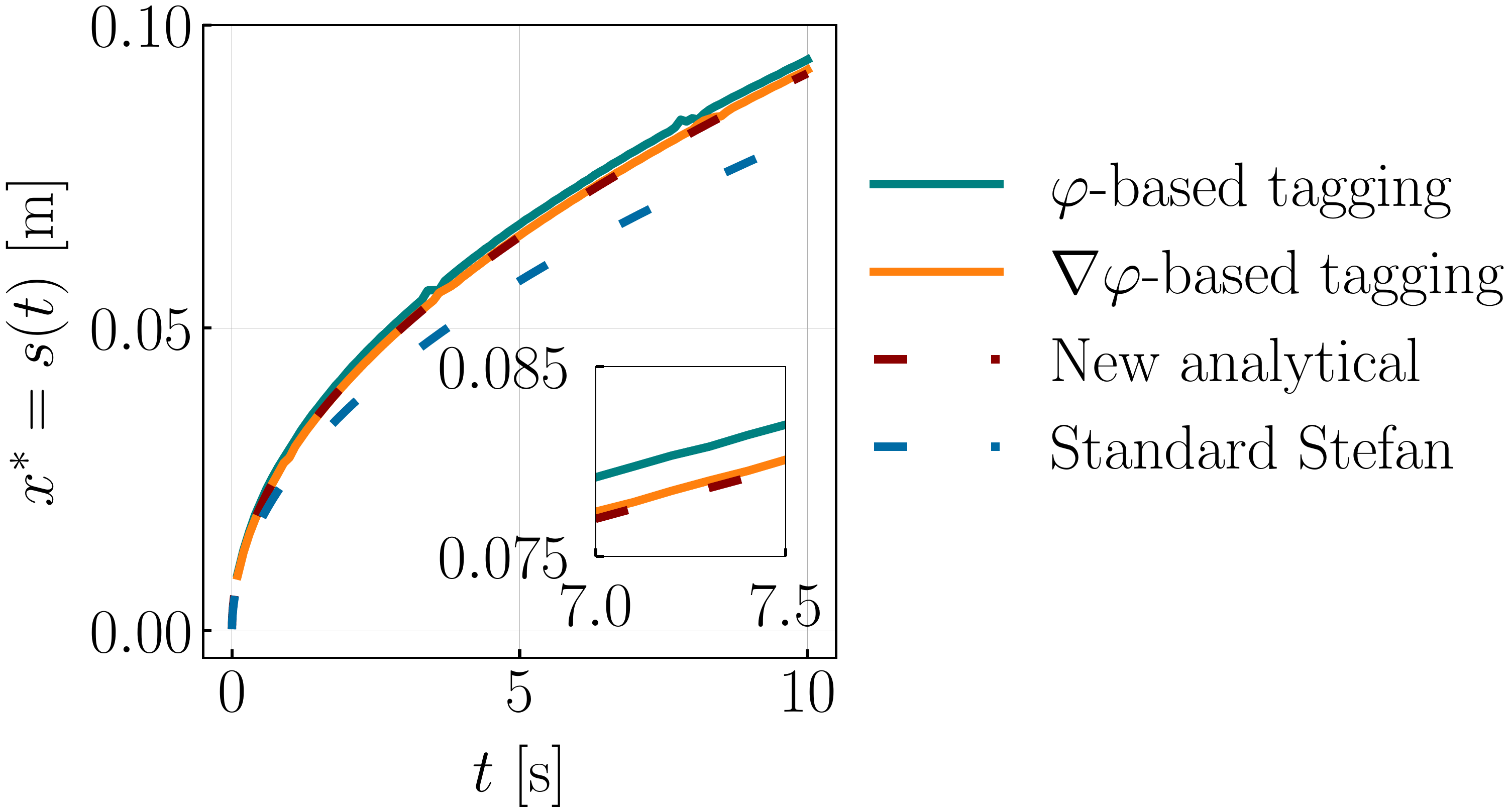}
\caption{Comparison of analytical and numerical interface positions obtained using $\varphi$-based and $\nabla \varphi$-based tagging criteria. Here, $R_\rho = \rhos/\rhol = 0.185$.}
\label{fig_lf_grad}
\end{figure}

Fig.~\ref{fig_lf_grad} shows a comparison between the time evolution of the solidification front and the analytical solution. In the figure the ``new analytical" solution refers to the Stefan problem that considers the volume expansion of PCM, whereas the ``standard Stefan" one does not. The next paragraph explains the difference between the two Stefan problems in more detail. As a result of uneven appearance and disappearance of the fine mesh level, $\varphi$-based tagging overpredicts the front position. Alternatively, $\grad \varphi$-based tagging produces more accurate results since the solidifying front is always resolved by finer meshes. Accordingly, we select the gradient-based tagging criteria to capture the moving phase change front in our simulations.

\vspace{1em}
\noindent \textbf{\underline{A word of caution for the reader:}}  In the literature, the solution to the standard Stefan problem is often confused with the solution that accounts for density-induced volume changes (which is derived in \ref{sec_2phase_analytical} and \cite{thirumalaisamy2023lowmach}). This confusion primarily stems from the way the temperature equation is written for the standard Stefan problem:
\begin{subequations} \label{eq_std_sp}
\begin{alignat}{2}
\rhos \cps  \frac{\partial \Ts}{\partial t}  &=\ks\frac{\partial^2 \Ts}{\partial x^2} \, \in  \Omegas(t),  \label{eq_temp_solid_ss}  \\
\rhol \cpl \frac{\partial \Tl}{\partial t} &=\kl\frac{\partial^2 \Tl}{\partial x^2} \,  \in \Omegal (t). \label{eq_temp_liquid_ss}
\end{alignat}
\end{subequations}
In the solid and liquid temperature equations, $\rhos$ and $\rhol$ can be different, so the solution to the standard Stefan problem can be misinterpreted as accounting for density-induced volume changes. Only one material property is needed in the heat equation: thermal diffusivity $\alpha = \kappa/(\rho C)$. Volume shrinkage/expansion effects are modeled by considering the convective term on the left side of Eqs.~\eqref{eq_std_sp}. Eqs.~\eqref{eq_nstd_sp} in \ref{sec_2phase_analytical} consider the convective term, whereas Eqs.~\eqref{eq_std_sp} do not. 

Bhattacharya~\cite{bhattacharya2019binary} presents a numerical method for simulating volume shrinkage during solidification. The author compared their numerical solution to the standard Stefan problem, assuming Eqs.~\eqref{eq_std_sp} account for volume shrinkage effects. Their numerical solution matches the analytical solution to the standard Stefan problem. It is a fortunate match. We demonstrated in our prior work that the solution to the 1D Stefan problem exhibiting shrinkage effects is quite similar to the standard Stefan problem. In contrast, the expansion problem shows a marked difference between the two solutions, which is apparent in Fig.~\ref{fig_lf_grad}.

\section{Complete solution algorithm} \label{sec_solution_algorithm}
Here we summarize the main steps involved in advancing the simulation for one time step $\Delta t = t^{n+1} - t^n$. A single cycle of fixed point iteration is considered.

\begin{enumerate}
    \item First, we advect the level set function $\REVIEW{\Phi}^n$ using the (non-divergence-free) velocity field $\mathbf{u}$ to obtain $\REVIEW{\Phi}^{n+1,k+1}$:
\begin{equation}
\frac{\REVIEW{\Phi}^{n+1,k+1} - \REVIEW{\Phi}^n}{\Delta t} + \left( \nabla \cdot (\REVIEW{\Phi} \mathbf{u}) \right)^{n+1,k} = (\REVIEW{\Phi} \nabla \cdot \mathbf{u})^{n+1,k}. \label{eq:phi_advect}
\end{equation}
The advected level set function is reinitialized to a signed distance function by solving the Hamiliton-Jacobi equation as proposed by Sussman et al.~\cite{Sussman1994}.

\item Similarly, we advect the smooth Heaviside function $H^n$ using the non-divergence-free velocity $\mathbf{u}$:
\begin{equation}
\frac{H^{n+1,k+1} - H^n}{\Delta t} + \left(\nabla \cdot (H \mathbf{u})\right)^{n+1,k} = (H \nabla \cdot \mathbf{u})^{n+1,k}. \label{eq_H_advect}
\end{equation}
Here, $H^n$ is computed from the level set function $\REVIEW{\Phi}^n$ using Eq.~\eqref{eq_smooth_H}. We remark that explicitly advecting $H$ is not essential; instead directly computing $H^{n+1,k+1}$ from $\REVIEW{\Phi}^{n+1,k+1}$ (through Eq.~\eqref{eq_smooth_H}) is sufficient. However, as demonstrated above for mass flux, the advective flux of Heaviside variable $H\u$ could be used to advect additional scalar variables of a more involved problem.

\item Next, we solve an additional mass balance equation at both cell- and face-centers of the mesh and compute the discrete mass flux $\V{m_\rho} = \rho \u$ in the process. The computed mass flux is used in the convective operator of the momentum equation ($\div (\rho \u \otimes \u)$ =  $\div (\V{m_\rho} \otimes \u)$), as well as in the advective operator of the enthalpy equation ($\div (\rho \u h) = \div (\V{m_\rho} h)$). 

The discrete mass balance equation reads as
\begin{equation}
\frac{\breve{\rho}^{n+1,k+1}-\rho^{n}}{\Delta t} + (\div  \V{m_\rho})^{n+1,k} = 0, \label{eq_mass_advect}
\end{equation}
which is solved to obtain the new density field $\breve{\rho}^{n+1,k+1}$ and the discrete mass flux $\V{m_\rho}$. This ensures the solver stability for high-density ratio flows. Additionally, the cell- and face-centered residuals of the mass balance Eq.~\eqref{eq_mass_advect} ($\mathcal{R}$ and $\V{\mathcal{R}}$, respectively) are computed. 

\item Using the newest density $\breve{\rho}$ and mass flux $\V{m_\rho}$, the nonlinear enthalpy equation is solved to update enthalpy $h$, temperature $T$, and liquid fraction $\varphi$.

The discrete enthalpy equation reads as
\begin{equation}
\frac{\breve{\rho}^{n+1,k+1} h^{n+1,k+1}-\rho^n h^{n}}{\Delta t} + (\div \V{m_\rho} h)^{n+1,k} = (\div\kappa\grad T)^{n+1,k+1} + \mathcal{R}h^{n+1,k} + Q_{\rm src}^{n+1,k}.
\label{eq_energy_discretized}
\end{equation}
The non-linear enthalpy equation is solved using Newton's method. See Sec.~\ref{sec_enthalpy_newton}.

\item Once the liquid fraction $\varphi^{n+1,k+1}$ is obtained by solving the enthalpy equation, it is extended from the PCM to the gas region by integrating Eq.~\eqref{eq_extrap} using forward Euler scheme as
\begin{equation}
\widetilde{\varphi}^{s+1} = \widetilde{\varphi}^{s} - \dt (1-H^{n+1, k+1}) \n^{n+1, k+1} \cdot \grad{\widetilde{\varphi}^{s}},
\end{equation}
with $s = 0,\ldots, 14$, and $\widetilde{\varphi}^{s = 0} = \varphi^{n+1, k+1}$. Eq.~\eqref{eq_extrap} is integrated for 15 time steps (with a CFL number of 0.3) so that $\varphi$ is sufficiently extended into the gas domain to calculate the surface tension force.


\item Finally, the low Mach Navier-Stokes equations are solved to obtain the new velocity $\u^{n+1,k+1}$ and pressure $p^{n+\frac{1}{2},k+1}$ fields 
\begin{subequations}
\begin{alignat}{2}
& \frac{\breve{\rho}^{n+1,k+1} \u^{n+1,k+1}-\rho^n \u^{n}}{\Delta t} + (\div [\V{m_\rho} \otimes \u])^{n+1,k}  = -\G p^{n+\frac{1}{2}, k+1} + (\L_\mu \u)^{n+\frac{1}{2}, k+1} +\breve{\vrho}^{n+1, k+1} \g   \nonumber  \\
&\hspace{18em} + \V{\mathcal{R}}\u^{n+1,k} - A_d^{n+1,k+1} \u^{n+1,k+1} + {\V{f}_{\rm st}}^{{\rm 3-phase}, n+\frac{1}{2},k+1},  \\
& \div{\u} = \begin{cases} 
0, & H < 0.5 \; (\text {i.e., in the gas phase}), \\
0,&  h<\hsol,\\ 
-\frac{ \rhos \rhol}{\rho^2}(\rhol-\rhos)H \frac{\displaystyle (\hliq-\hsol)}{\displaystyle \left(h(\rhol- \rhos)- \rhol \hliq +\rhos \hsol \right)^2}\displaystyle \left(\div {\kappa\grad{T}} + Q_\text{src}  \right),&  \hsol \le h \le \hliq, \\ 
0,&  h>\hliq.\\ 
\end{cases} 
\end{alignat}
\end{subequations}
In our solution methodology, velocity and pressure are solved simultaneously through the use of a matrix-free flexible GMRES (FGMRES) solver. A projection method-based preconditioner is used that treats the stiff Carman-Kozeny drag term $-A_d \u$ implicitly. More details on the low Mach Navier-Stokes solver and projection method-based preconditioner can be found in our previous work~\cite{thirumalaisamy2023pre}.    

\end{enumerate}

\section{Software implementation}
The low Mach enthalpy method and its proposed improvements described in this work have been implemented within the IBAMR library \cite{IBAMR-web-page}, an open-source C++ software enabling the simulation of CFD and fluid-structure interaction problems. IBAMR utilizes adaptive mesh refinement (AMR) and Cartesian grid management through the SAMRAI framework \cite{HornungKohn02, samrai-web-page}, with linear solver support provided by the PETSc library (\cite{petsc-user-ref, petsc-web-page}). \REVIEW{All simulations presented in this work are implemented in the following GitHub repository:\url{https://github.com/IBAMR/IBAMR/tree/phase_change_patch_6}.}
 
 \section{Results and discussion}  \label{sec_results}
 
\subsection{Isothermal advection of a \REVIEW{droplet}}  \label{sec_bubble advection_enthalpy}

To demonstrate the combined consistency of mass, momentum, and enthalpy integrators, we modify the \REVIEW{droplet} advection problem discussed in Sec.~\ref{sec_temp_discretization}: in addition to mass and momentum transport, we also consider enthalpy transported by the \REVIEW{droplet}\footnote{\REVIEW{This benchmark test is provided in IBAMR GitHub within the directory \texttt{examples/phase\_change/ex2}.}}. Thermal conductivity $\kappa$ and $Q_\text{src}$ are set to zero. All other problem parameters (e.g., \REVIEW{droplet} size, boundary conditions, etc.) remain the same as in  Sec.~\ref{sec_temp_discretization}. In the absence of thermal diffusion and body forces, the system of equations reduces to
\begin{subequations} \label{eq_isothermbubble}
 \begin{alignat}{2}
& \D{\REVIEW{\Phi}}{t}+\u\cdot\grad{\REVIEW{\Phi}}=0, \\
& \D{\rho}{t}+\div{\left(\rho \u\right)}=0, \\
& \D{\rho h}{t}+\div{\left(\rho \u h\right)}=0, \\
& \D{\left(\rho \u\right)}{t}+\div{\left(\rho \u \otimes \u\right)} = 0.
\end{alignat}
\end{subequations}

The initial temperature of the liquid \REVIEW{droplet} and the surrounding gas is set to $T_i = 3$ K~\footnote{For this test problem, units of physical quantities are not relevant (as there are no right-hand side terms in Eqs.~\eqref{eq_isothermbubble}), but are included to maintain notational consistency.} and the fusion/phase change temperature of the liquid \REVIEW{droplet} is set to $T_m = 2$ K. The liquidus and solidus temperatures are $\Tliq = 2.1$ K and $\Tsol = 1.9$ K, and a latent heat value of $L = 100$ J/kg is used. The specific heats of liquid and gas phases are 1042.4 J/kg $\cdot$ K and 1000 J/kg $\cdot$ K, respectively. For this problem, we expect that: 
\begin{itemize}
\item In the absence of body forces (pressure and viscous) the velocity should remain constant throughout the simulation.
\item The \REVIEW{droplet} gets advected with constant velocity, retains its shape, and returns to its original position after each time period.
\item With no energy sources and non-conducting fluids, enthalpy and temperature should remain constant.
\item There should be no phase change or spurious phase generation. This also means that specifying latent heat value is arbitrary for this problem. 
\item The total mass, momentum, and enthalpy in the domain should remain constant.
\end{itemize}

\begin{figure}
\centering
\includegraphics[width=1.0\linewidth]{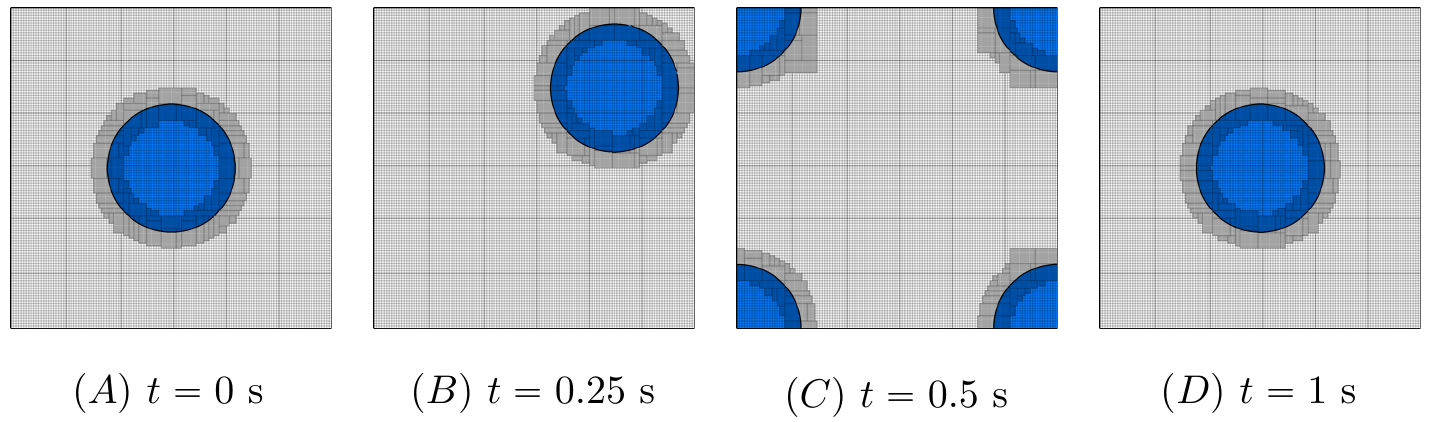}
\caption{Advection of a \REVIEW{droplet} with density ratio $\rho_i/\rho_o = 10,000$ in an initially uniform velocity field $\u = (u,v) = (1,1)$. The simulation is performed using two levels of mesh refinement. The coarse grid size is $N_x \times N_y = 128^2$. A unit periodic domain is considered. $\kappa$, $\mu$ and $Q_\text{src}$ are set to zero.}
\label{fig_bubble_adv_dynamics}
\end{figure}

Fig.~\ref{fig_bubble_adv_dynamics} illustrates the advection of the \REVIEW{droplet} at various time instances. The simulation employs $l = 2$ levels of mesh with the coarse grid size set to $N_{x0} \times N_{y0} = 128^2$, and a refinement ratio of $\nref = 2$. The simulation runs until $t=10$ s. As shown in Fig.~\ref{fig_bubble_adv_dynamics}, the \REVIEW{droplet} moves with a constant velocity and returns to its original position after one time period $\tau = \frac{d}{|\u_c|} = 1$ , in which $d = \sqrt{2}$ and $|\u_c| = \sqrt{2}$. Additionally, the \REVIEW{droplet} undergoes minimal distortion while returning to its original position. Throughout the simulation (data not shown here for brevity), velocity remains constant at its initial value $\u_c = (u,v) = (1,1)$. In addition, the enthalpy equation solution does not result in spurious phases (for example, liquid becoming solid). 

\begin{figure}
\centering
\includegraphics[width=1.0\linewidth]{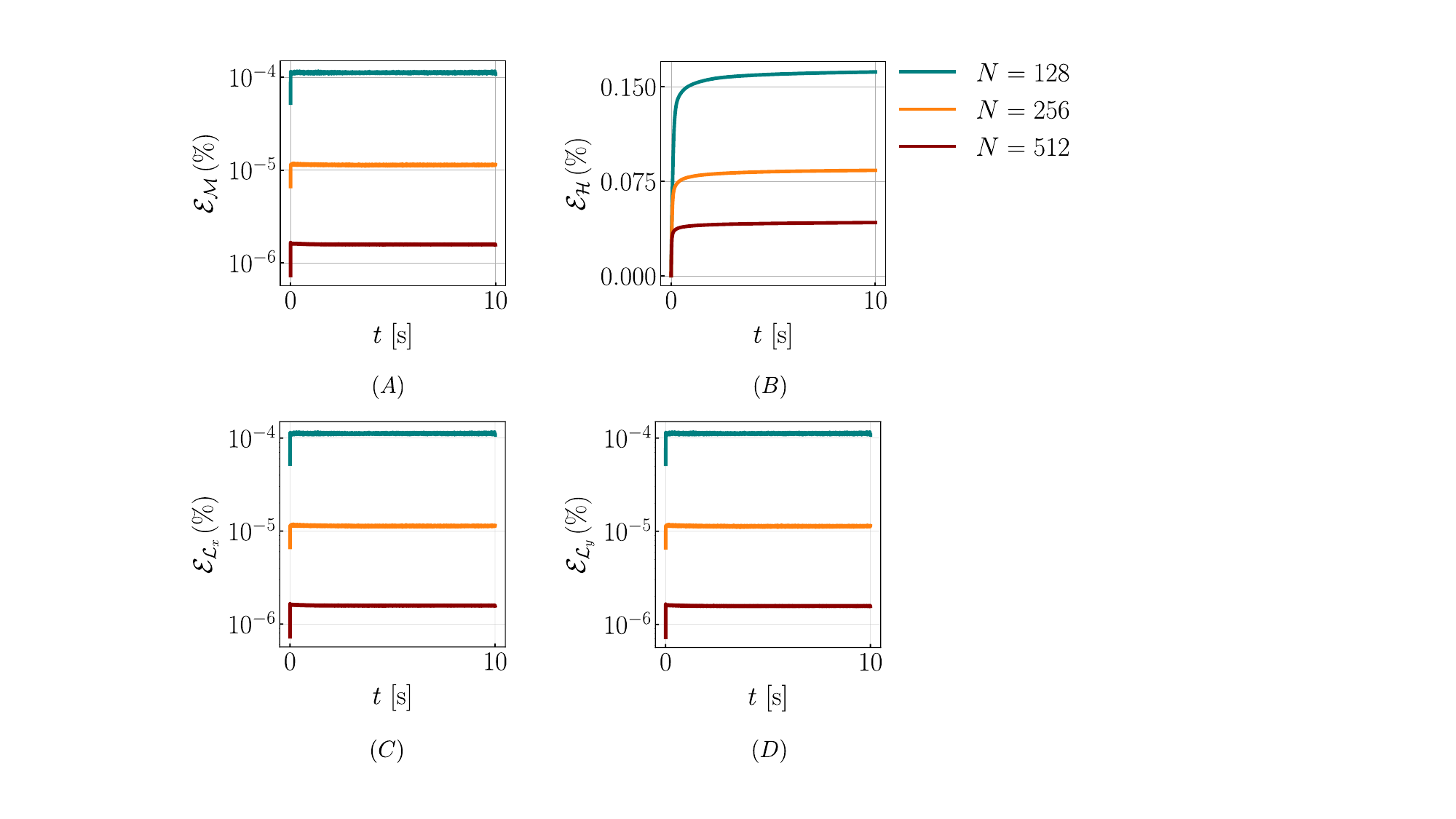}
\caption{\REVIEW{Grid convergence study of system's mass $\mathcal{M} =  \int_\Omega \rho\, \rm{d}V$, $x$-momentum $\mathcal{L}_{\x} =  \int_\Omega \rho u\, \rm{d} V$,  $y$-momentum $\mathcal{L}_{\x} =  \int_\Omega \rho v\, \rm{d} V$ and enthalpy $\mathcal{H} =  \int_\Omega \rho h\, \rm{d} V$. }}
\label{fig_bubble_adv_convergence}
\end{figure}

\begin{figure}
\centering
\includegraphics[width=1.0\linewidth]{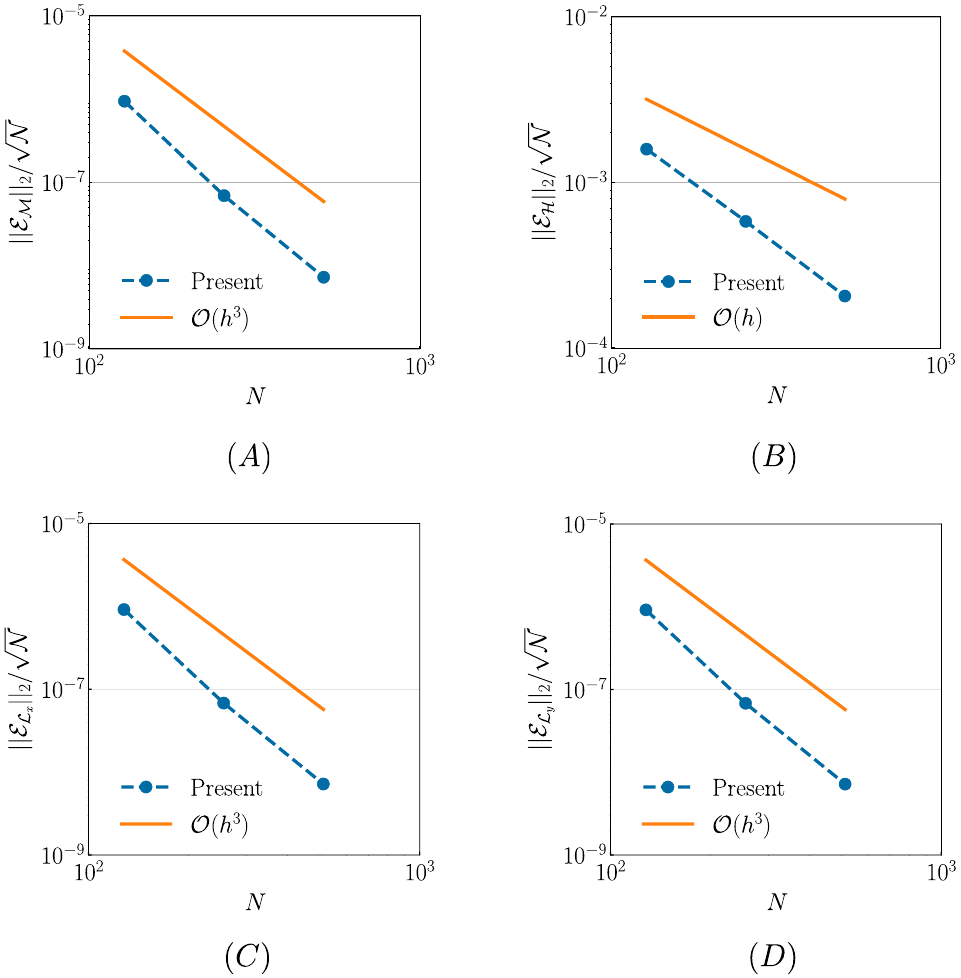}
\caption{\REVIEW{Convergence rates of the low Mach enthalpy method considering the isothermal advection of a \REVIEW{droplet}. The $\mathcal{L}^2$ error for a quantity $\psi$ is defined to be the root mean squared error (RMSE) of the vector
 $||\mathcal{E}_\psi ||_{\rm RMSE} =||\mathcal{E}_\psi||_2/\sqrt{\mathcal{N}}$, in which $\mathcal{N}$ denotes the size of the vector $\mathcal{E}_\psi$. Here, $\Psi$ represents mass $\mathcal{M} =  \int_\Omega \rho\, \rm{d}V$, $x$-momentum $\mathcal{L}_{\x} =  \int_\Omega \rho u\, \rm{d} V$,  $y$-momentum $\mathcal{L}_{\x} =  \int_\Omega \rho v\, \rm{d} V$ and enthalpy $\mathcal{H} =  \int_\Omega \rho h\, \rm{d} V$.} }
\label{fig_bubble_adv_ooa}
\end{figure}

To quantify the accuracy of consistent mass-momentum-enthalpy integrators, a grid convergence study is conducted. We consider uniform grids of size $N_x \times N_y = \{128^2, 256^2, 512^2\}$ for the convergence study. This is to avoid discretization errors near the coarse-fine boundary of adaptively refined grids. Errors in total mass $\mathcal{M} =  \int_\Omega \rho\, \rm{d} V$, $x$-momentum $\mathcal{L}_{x} =  \int_\Omega \rho u\, \rm{d} V$,  $y$-momentum $\mathcal{L}_{y} =  \int_\Omega \rho v\, \rm{d} V$ and enthalpy $\mathcal{H} =  \int_\Omega \rho h\, \rm{d} V$ are computed using Eq.~\eqref{eq_rel_change}, and the results are shown in Fig.~\ref{fig_bubble_adv_convergence}. The system's mass, momentum, and enthalpy remain constant over time, with the exception of the initial transient phase where the initially sharp interface diffuses. Additionally, the temperature remains mostly constant throughout the simulation ($T \approx 2$ K) except near the interface (data not shown for brevity). This is because the primary variable $h$ diffuses over time and $T$ is reconstructed based on $T$-$h$ relations in our solution algorithm. The temperature change near the interface does not induce phase change, however. \REVIEW{As shown in the Fig.~\ref{fig_bubble_adv_ooa}, the mass and momentum transport is third-order accurate ($\mathcal{O}(\Delta x^3)$) using this method}. As velocity remains constant (numerically), mass and momentum errors are the same. However, enthalpy solution is only first-order ($\mathcal{O}(\Delta x)$) accurate. The behavior of error as a function of grid resolution can be explained as follows. At the end of each time step, the density field is synchronized with the level set function. This limits density (and momentum, in this case) diffusion due to advection. The level set function is reinitialized to a signed distance function using a third-order accurate ENO scheme; see our prior works on level set reinitialization~\cite{nangia2019robust}. Enthalpy transport, however, is subject to the upwinding errors of the CUI limiter, which displays first-order accuracy near non-monotone regions. These are regions close to the liquid-gas interface. The CUI scheme's first-order accuracy corresponds to Gudonov's order barrier theorem.

These results indicate that our mass, momentum, and enthalpy integrators are capable of stably simulating high density ratio flows without generating spurious momentum or phases. 



\subsection{Thermocapillary flows}  \label{sec_thermo_capillary_flow}

Here we solve enthapy Eq.~\eqref{eq_enthalpy} in conjunction with surface tension Eq.~\eqref{eq_fst_d} to simulate surface tension-driven flows. Specifically, the thermocapillary migration of gas bubbles embedded in liquid (PCM) is considered. In most thermocapillary flow models in the literature, the temperature equation (without consideration to any phase change) is used instead of the enthalpy equation \cite{nas2003thermocapillary, herrmann2008thermocapillary, ma2011direct, seric2018direct, tripathi2018motion}. Using an enthalpy equation instead of temperature does not matter on a continuous level, but discretely, it may. This test aims to determine whether the low Mach enthalpy integrator is capable of accurately simulating thermocapillary flows without inducing spurious phase changes. 

\begin{figure}
\centering
\includegraphics[width=0.45\linewidth]{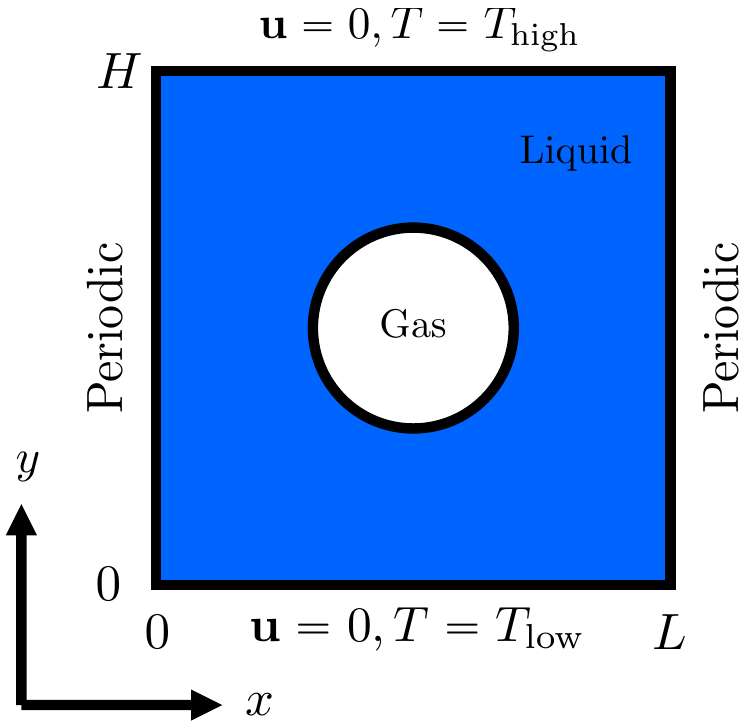}
\caption{Schematic of the thermocapillary migration of a gas bubble in 2D. A gas bubble of radius $R$ rises in an ambient fluid due to a gradient in the surface tension coefficient. The temperature varies linearly $T_{\rm low} \le T \le T_{\rm high}$ with bottom and top wall temperatures of $T_{\rm low}$ and $T_{\rm high}$, respectively. No-slip boundary conditions are used for velocity at the top and bottom walls. Periodic boundary conditions are considered for vertical walls.}
\label{fig_marangoni_schematic}
\end{figure}

Consider a computational domain $\Omega \in [0, L]^d$, $d = 2$ and 3, within which a bubble of radius $R$ is immersed in the ambient fluid, as illustrated in Fig.~\ref{fig_marangoni_schematic}. The bubble's initial centroid is in the middle of the domain. The ratio of gas bubble thermophysical properties ($\mu$, $\rho$, $C$, $\kappa$) to the surrounding fluid is denoted $\gamma$. The top and bottom boundaries are considered walls with no-slip (zero-velocity) boundary conditions. Temperature varies linearly in the domain $T_{\rm low} \le T \le T_{\rm high}$. Here, $T_{\rm low}$ and $T_{\rm high}$ represent temperatures at the bottom and top walls, respectively. All other boundaries are periodic. 


The surface tension coefficient varies with temperature as 
\begin{equation}
\sigma = \sigma_0 + \frac{\rm{d} \sigma}{\rm{d} T}\bigg|_{T_0}(T - T_0),
\label{eq_sigma}
\end{equation}
in which $ \sigma_0$ and $\frac{\rm{d} \sigma}{\rm{d} T}\big|_{T_0}$ are the surface tension and Marangoni coefficients (respectively) computed at the reference temperature $T_0$. The gas bubble moves upward due to the variation in surface tension caused by the imposed temperature gradient. Key non-dimensional parameters of the problem are the Reynolds number $\displaystyle{Re = \frac{ \rhol V R}{ \mul}}$, Marangoni number $\displaystyle{Ma= \frac{ V R }{ \alphal}}$, Capillary number $\displaystyle{Ca=\frac{V \mul}{\sigma_0}}$, and Prandtl number $\displaystyle{Pr=\frac{\mul \cpl}{\kl}}$. Here, $V$ is the velocity scale defined as $\displaystyle{V = \bigg| \frac{ \frac{\rm{d} \sigma}{\rm{d} T}\big|_{T_0}  \nabla T R}{\mul}} \bigg|$. Two different cases are considered: zero Marangoni and finite Marangoni. The zero Marangoni case checks the accuracy of the surface tension force model\footnote{\REVIEW{This benchmark test is provided in IBAMR GitHub within the directory \texttt{examples/multiphase\_flow/ex14}.}}. The finite Marangoni case examines the low Mach enthalpy integrator's ability to simulate thermocapillary flows\footnote{\REVIEW{This benchmark test is provided in IBAMR GitHub within the directory \texttt{examples/phase\_change/ex7}.}}. 

\subsubsection{Zero Marangoni number}
A time-invariant linear temperature field ($\kappa \rightarrow \infty$) is considered, which is referred to as the ``zero Marangoni number" problem in the literature. Young et al.~\cite{young1959motion} investigated the zero Marangoni number thermocapillary migration of a bubble in ambient fluid. The authors provide an analytical expression for the terminal velocity of the bubble which reads as 
\begin{equation}
V_t= -\frac{2  \frac{\rm{d} \sigma}{\rm{d} T}\big|_{T_0} \nabla T R}{6\mul+9\mug}
\label{eq_young_rise_vel}
\end{equation}
We compare the numerically computed bubble rise velocity with the analytical solution to assess the surface tension force model accuracy. The normalized bubble rise velocity is numerically calculated as
\begin{equation}
V_{\rm rise} = \frac{\int_{\Omega} v(1-H) \text{d} V}{V_t \int_{\Omega} (1-H) \d V}.
\end{equation}

\begin{figure}
\centering
\includegraphics[width=1.0\linewidth]{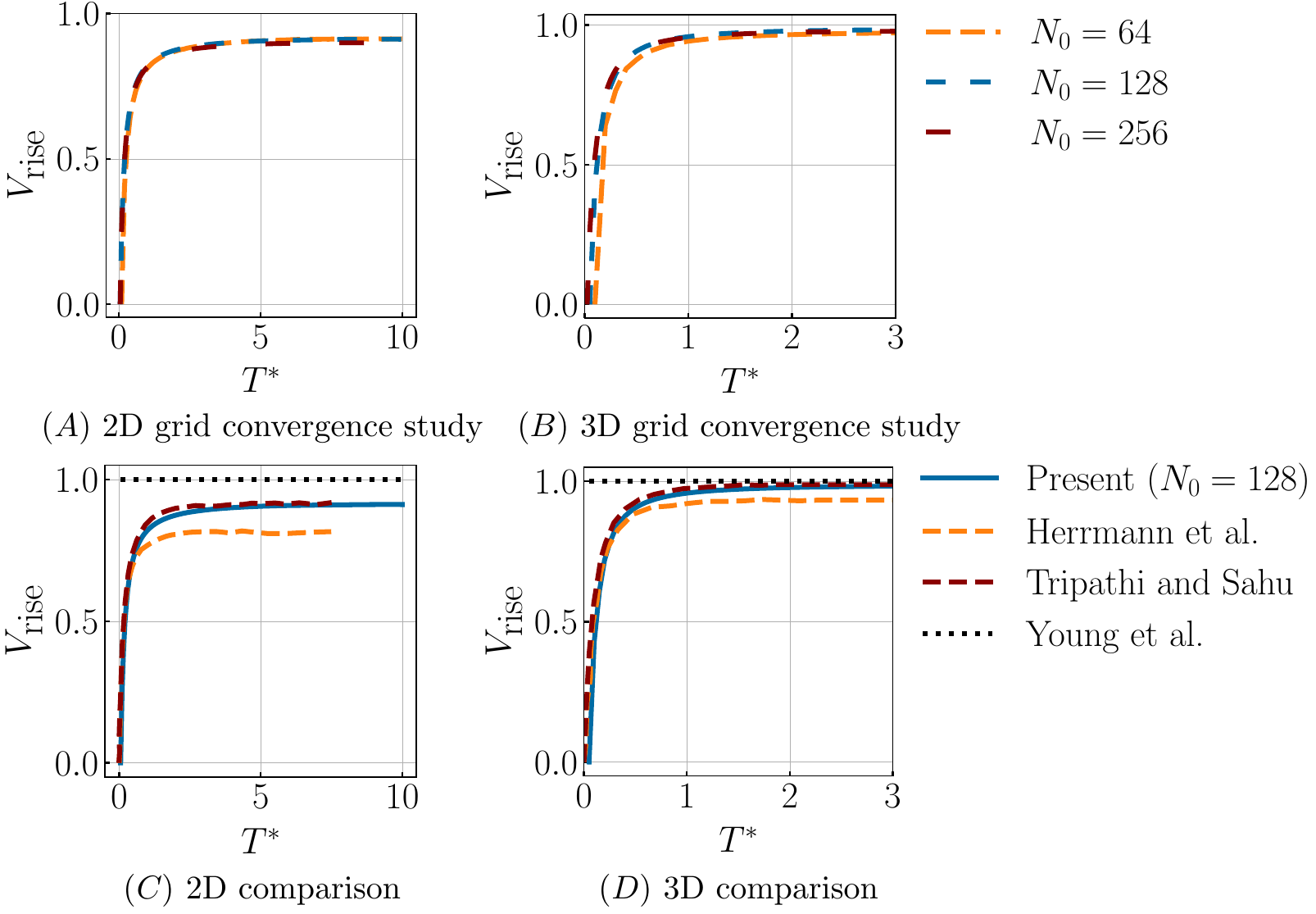}
\caption{Normalized bubble rise velocity $V_{\rm rise}$ as a function of normalized time $T^* = tV/R $ for the case of $Ma = 0$: Grid convergence study for (A) 2D  and (B) 3D simulations. Comparison with other studies in (C) 2D and (D) 3D. }
\label{fig_zero_ma_result}
\end{figure}

Due to the time-invariant nature of the linear temperature field, we do not solve the enthalpy equation, since it would change the temperature field numerically, even slightly. The radius of the bubble is taken to be $R = 0.5$ m. The ratio of thermophysical properties of the gas and ambient fluid is set to $\gamma = 1$. In this case, the density and viscosity of the bubble (and also the liquid) are 0.2 kg/m$^3$ and 0.1 kg/m$\cdot$s. The square and cubic computational domains are of length $L=15R$ in 2D and 3D, respectively. The surface tension parameters are $ \sigma_0 = 0.1$ N/m and $\frac{\rm{d} \sigma}{\rm{d} T}\big|_{T_0} = -0.1$ N/m$\cdot$K. Temperatures at the bottom and top walls are  $T_\text{low} = 0$ K and $T_\text{high} = 1$ K.  This corresponds to a temperature gradient of $\displaystyle{\D{T}{y}=0.133}$ K/m. These dimensional values give us $V = 6.67\times 10^{-2}$ m/s, $Re = Ca = 6.67\times 10^{-2}$ and $Ma = Pr = 0$.  Grid convergence studies are performed using three grid resolutions: coarse, medium and fine. These grids have a coarse level ($l = 1$) resolution of $N_0$ = 64, 128 and 256, respectively. Two levels of grid refinement with a refinement ratio of $\nref=2$ are employed for all three grids. A uniform time step size of $\Delta t = 0.1$ s is used for the coarse grid. $\Delta t$ is halved for each successively refined mesh. The top row of Fig.~\ref{fig_zero_ma_result} shows the time evolution of the normalized rise velocity plotted against normalized time $T^* = tV/R $ for the three grids in both two and three spatial dimensions. Convergent behavior for rise velocity can be observed in both 2D and 3D simulations. Consequently, we choose the medium grid (with $N_0 = 128$) for comparing our results with prior studies~\cite{herrmann2008thermocapillary, young1959motion,tripathi2018motion}. \REVIEW{From Figs.~\ref{fig_zero_ma_result} (C) and (D), it can be observed that our results are in agreement with those of Tripathi and Sahu~\cite{tripathi2018motion} for both 2D and 3D simulations. Our 2D simulation deviates from the theoretical study of Young et al.~\cite{young1959motion} by 8\%, whereas the 3D simulation shows a deviation of 1.5\%. In comparison, Herrmann et al.'s~\cite{herrmann2008thermocapillary} work exhibits deviations of approximately 18\% and 6.3\% from the theoretical rise velocity for 2D and 3D simulations, respectively.}

Results from the zero Marangoni case validate the accuracy of the surface tension force model as presented in Eq.~\eqref{eq_fst_d}. However, it does not confirm the accuracy of the low Mach enthalpy integrator for simulating thermocapillary flows. Next, a finite Marangoni case is considered to verify this.  


 \subsubsection{Finite Marangoni number}

This section investigates the thermocapillary migration of gas bubbles at a finite Marangoni number, $Ma = 0.72$. Other works~\cite{nas2003thermocapillary,ma2011direct, seric2018direct, tripathi2018motion} have also considered the $Ma = 0.72$ case for validating multiphase flow solvers. As temperature varies with time (and space), the Navier-Stokes and enthalpy equations are coupled through the advection term in the enthalpy equation and the surface tension force in the momentum equation. 
 
The gas bubble in the simulations has a radius of $R = 1.44\times10^{-3}$ m and is initially positioned at the center of the computational domain. A square and a cubic computational domain of length $L=4R$ are used in 2D and 3D simulations, respectively. The thermophysical properties of the ambient liquid are $\rhol = 500$ kg/m$^3$, $\mul = 0.024$ kg/m$\cdot$s, $\kl = 2.4\times10^{-6}$ W/m$\cdot$K, and $\cpl = 10^{-4}$ J/kg$\cdot$K. The ratio of the thermophysical properties of the gas bubble with the surrounding fluid is set to $\gamma=0.5$. The surface tension parameters are $ \sigma_0 = 10^{-2}$ N/m, $\frac{\rm{d} \sigma}{\rm{d} T}\big|_{T_0} = -2\times10^{-3}$ N/m$\cdot$K. Temperatures at the bottom and top walls are  $T_\text{low} = 289.424$ K and $T_\text{high} = 290.576$ K. This corresponds to a temperature gradient of $\displaystyle{\D{T}{y}=200}$ K/m. The phase change/fusion temperature of the ambient liquid is set to $T_m$ = 265 K, with $\Tliq = 270$ K, $\Tsol = 260$ K, $T_{\rm ref} = 290$ K. A consistent enthalpy solver should not induce solidification of the liquid since the imposed wall temperatures are higher than the fusion temperature.  According to the problem parameters, $Re=Ma=0.72$, $Ca=0.0576$, $V = 0.024$ m/s and $Pr = 1$ for this problem.
 
The simulations are run until $t = 0.12$ s. Grid convergence studies are performed in both 2D and 3D. The 2D grids considered have coarse level resolutions of $N_0 = 64, 128$, and $256$ with two levels of mesh refinement and $\nref=2$. A uniform time step size of $\Delta t = 2.5\times10^{-5}$ s is used for the coarse mesh ($N_0 = 64$). This choice of $\Delta t$ respects the explicit capillary time step size restriction~\cite{brackbill1992continuum, denner2022breaching}, which is given by $\Delta t_\sigma = \sqrt{\frac{\rhol+\rhog}{4\pi\sigma} \Delta x^3} \approx 2.33 \times 10^{-5}$ s. With each successive grid refinement (by a factor of 2), the time step size is reduced by a factor of 4~\footnote{\REVIEW{Compared to the theoretical formula, which would reduce $\Delta t$ by a factor of about $\sqrt{2^3}\approx 3$, a reduction of 4 seems more conservative. In practice, a large variation in the curvature of the interface further reduces $\Delta t$.}}. For 3D simulations, coarse level grids of size $N_0 = 32, 64$, and $128$ are considered, with a uniform time step size of $\Delta t = 2 \times 10^{-5}$ s for the coarse mesh. $\Delta t$ is halved for each fine grid, which maintains the explicit capillary time step size restriction.
 
\begin{figure}
\centering
\includegraphics[width=1.0\linewidth]{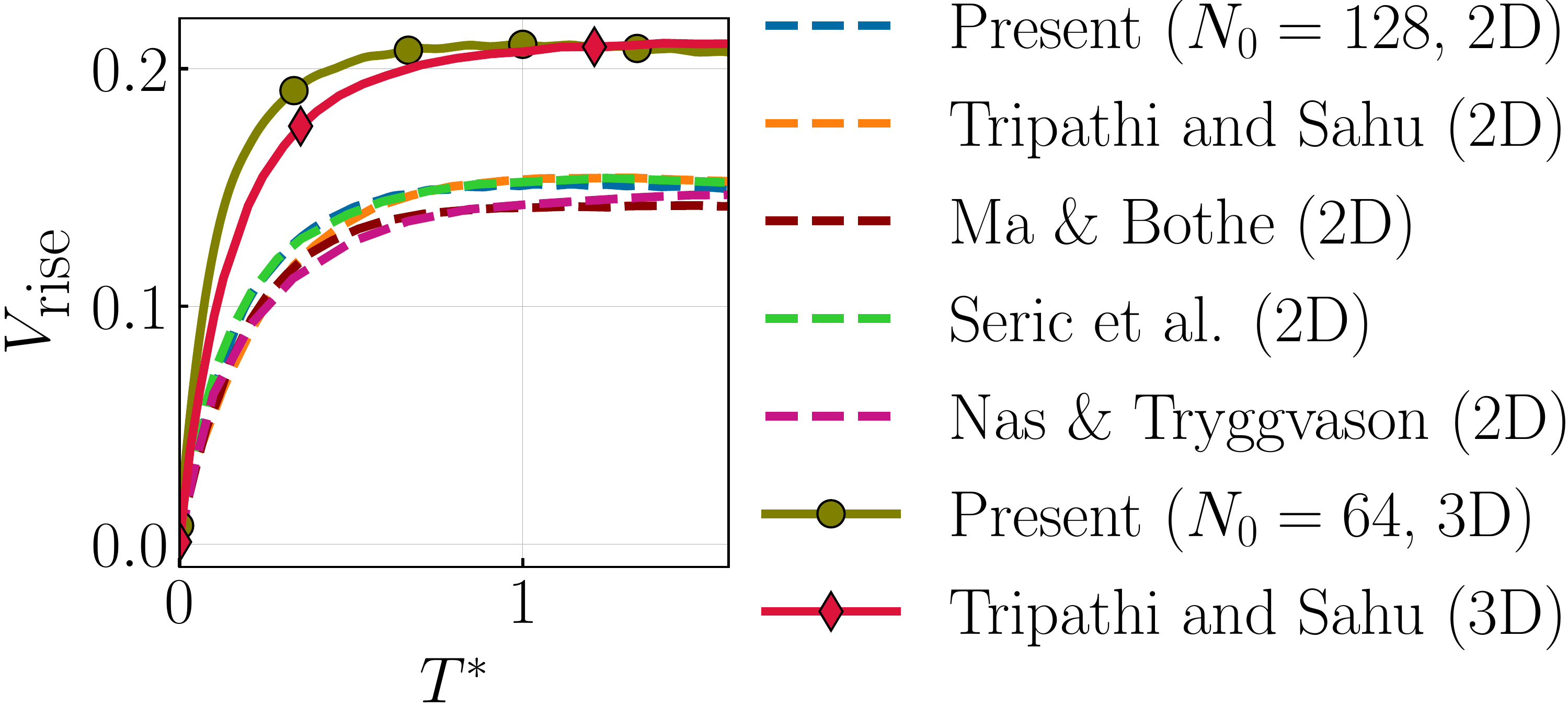}
\caption{\REVIEW{Normalized bubble rise velocity $V_{\rm rise}$ as a function of normalized time $T^* = tV/R $ for the case of $Ma = 0.72$: Grid convergence study for (A) 2D and  (B) 3D simulations. (C) Comparison of 2D and 3D results with prior works.}}
\label{fig_finite_ma_result}
\end{figure}

The results are presented in Fig.~\ref{fig_finite_ma_result} (A) for 2D simulations and Fig.~\ref{fig_finite_ma_result} (B) for 3D simulations. Both 2D and 3D simulations show consistent bubble rise velocity across all grid resolutions. Using the medium grid, we compare our results with previous studies. A comparison of the results is shown in Fig.~\ref{fig_finite_ma_result} (C). Our results match closely with those obtained by Tripathi and Sahu~\cite{tripathi2018motion}, who used the volume of fluid method within the open-source~\texttt{Basilisk} code. Furthermore, the authors in~\cite{tripathi2018motion} solved the temperature equation instead of the enthalpy equation.   

The results of this section suggest that the low Mach enthalpy method (which also handles phase change) is able to simulate thermocapillary flows accurately. This capability is attributed to the consistency of mass, momentum and enthalpy integrators that do not result in spurious mass, momentum, enthalpy or phase generation.  

\subsection{Stefan problem with melting-induced volume changes}  \label{sec_stefan_prob_melting}

In this section, we present a simple problem that can be used to validate advanced CFD codes that model metal manufacturing processes like welding and metal 3D printing. Numerical studies have relied on experimental data to validate heat sources (e.g., laser beams) induced melting of metals and alloys. Due to the high number of design parameters involved in experiments (shielding/inert gas flow to prevent oxidation, laser beam reflection, etc.) and uncertainties (for example, the conductivity of metals and their alloys may not be known fully) that cannot be fully accounted for in simulations, previous studies have altered the thermophysical properties of the PCM artificially to match the results.  

In our previous work~\cite{thirumalaisamy2023lowmach}, we provided an analytical solution to the two-phase Stefan problem involving solidification of a PCM exhibiting a volume change during the process. Building upon that, we derive an analytical solution (see \ref{sec_2phase_analytical}) to the two-phase melting problem with volume changes. We deduce two additional cases, useful for code validation, from the Stefan model problem: 
\begin{itemize}
\item Two-phase melting problem with a flux boundary condition.
\item Three-phase melting problem with a passive gas phase and a volumetric heat source in the domain.
\end{itemize}

The modeled PCM's thermophysical properties are largely iron-based, as shown in Table~\ref{tab_melting_properties}. Solid phase density is $\rhos =  8100$ kg/m$^3$, and liquid phase density is $\rhol = 7165.384$ kg/m$^3$. As the liquid phase has a lower density than the solid phase, PCM expands upon melting.

 \subsubsection{Two-phase melting problem with flux boundary condition}  \label{sec_2phase_stefan_prob_melting}
\begin{figure}
\centering
\includegraphics[width=1.1\linewidth]{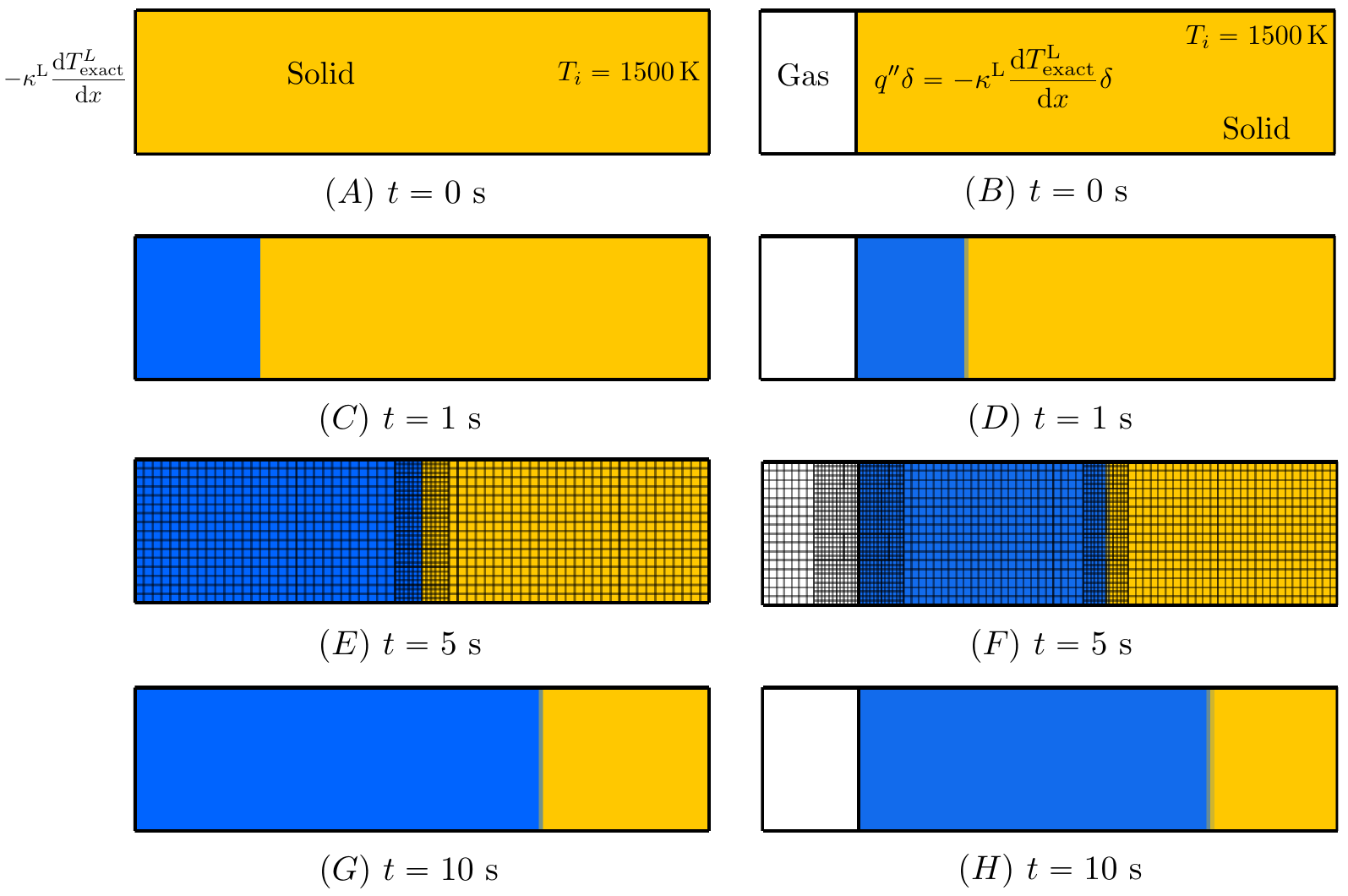}
\caption{Stefan problem with melting: time evolution of the liquid fraction $\varphi$ and adaptive grids. For the left column plots, melting occurs due to the flux boundary condition at the left end of the domain. For the right column plots, flux boundary conditions are imposed through a volumetric term in the enthalpy equation. There is also a passive gas phase. Gradient-based tagging is used to generate AMR grids. Solid, liquid and gas regions are depicted in yellow, blue and white colors. \REVIEW{The AMR grids are shown only at $t=5$ s.}}
\label{fig_stefan_melting_dynamics}
\end{figure}

 Consider a quasi one-dimensional computational domain $\Omega \in [0,0.1] \times [0, 0.005]$ as illustrated in Fig.~\ref{fig_stefan_melting_dynamics}(A)\footnote{\REVIEW{This benchmark test is provided in IBAMR GitHub within the directory \texttt{examples/phase\_change/ex9}.}}. The simulation employs a two-level mesh with coarse cell size of $\Delta x= \Delta y = 0.005/32$.  The simulation runs until $t=10$ s with a constant time step size of $\Delta t = 10^{-3}$ s. Initially, the domain is fully occupied by the solid PCM with a temperature of $T_i=1500$ K. At the right boundary ($x=0.1$), a zero pressure (outflow) condition is used for the flow solver and an adiabatic (homogeneous Neumann) condition is used for the temperature $T$. At the left end of the domain ($x=0$), no-slip and inhomogeneous Neumann $\displaystyle{\left. -\kl \frac{{\rm d} \Tl_\text{exact}}{{\rm d} x} \right|_{x=0}}$ boundary conditions are employed for velocity and temperature, respectively. Here, $\Tl_\text{exact}$ denotes the analytical temperature of the liquid phase as derived in~\ref{sec_2phase_analytical}. The top and bottom boundaries (in the $y$-direction) are periodic.
 
\begin{figure}
\centering
\includegraphics[width=1.0\linewidth]{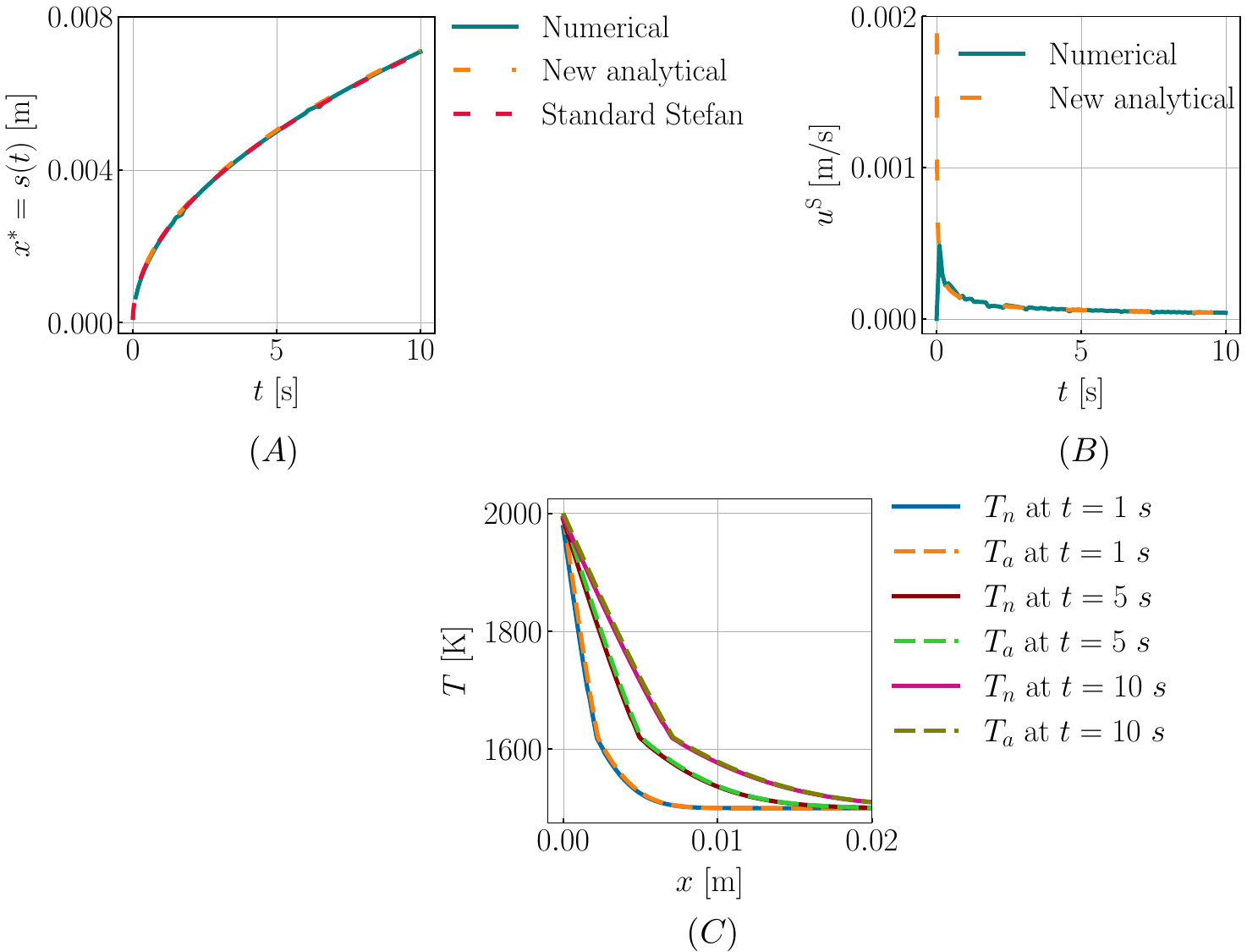}
\caption{Comparison of CFD and analytical solutions for the two phase melting problem with flux boundary conditions: (A) interface position; (B) solid velocity; and (C) temperature.}
\label{fig_stefan_melting_without_gas}
\end{figure}

 The schematic of the problem is depicted in Fig.~\ref{fig_stefan_melting_dynamics}(A). Since $R_\rho = \rhos/\rhol > 1$, there is a volume expansion upon melting, which induces additional flow. As shown in~\ref{sec_2phase_analytical}, the solid phase moves with a rigid body velocity $\us (t)$, while the liquid velocity $\ul$ remains zero. To simulate this behavior, the solid fraction $\varphi_{\rm S}$ in the Carman-Kozeny drag coefficient is replaced with the liquid fraction $\varphi$. Consequently, the Carman-Kozeny drag force retards liquid phase motion. Gradient-based tagging is used to generate adaptively refined grids. Melting initiates due to the flux boundary condition and the liquid-solid interface moves rightwards over time until the PCM completely melts. The evolution of the liquid fraction and the adaptive mesh is shown in the left column of Fig.~\ref{fig_stefan_melting_dynamics}. We compare the numerical solutions (interface position $x* = s(t)$, velocity of the solid $\us(t)$, temperature distribution $T(x)$) with the analytical ones in Fig.~\ref{fig_stefan_melting_without_gas}. An excellent match is observed between the two. \REVIEW{As shown in Fig.~\ref{fig_stefan_melting_dynamics}(B), a large spike is observed in the solid velocity at the initial time. This occurs because the interface velocity $u^* = \frac{{\rm d} s }{{\rm d} t} \rightarrow \infty$ (see Appendix~A) at time $t= 0^+$. Since the solid velocity is proportional to the interface speed, the CFD simulation produces a large solid velocity at the beginning.} The new analytical and standard Stefan problem solutions are very similar for this problem. This is attributed to the 1D nature of the problem as well as the low density difference between the solid and liquid phases ($\approx 11\%$). Nevertheless, this amount of density variation may cause PCM free surfaces to rise or protrude in 2D or 3D problems~\cite{thirumalaisamy2023lowmach}. 

\begin{table}[]
\centering
\caption{Thermophysical properties of the PCM (largely iron-based) used to simulate the melting problem.}
\label{tab_melting_properties}
\begin{tabular}{ll}
Property & Value \\
\midrule
Thermal conductivity of solid   $\ks$     &   22.9  W/m.K  \\
Thermal conductivity of liquid  $\kl$                 & 22.9  W/m.K    \\
Specific heat of solid  $\cps$                 & 627 J/kg.K      \\
Specific heat of liquid  $\cpl$                  &  723.14  J/kg.K   \\
Solidification temperature  $T_m$                &  1620 K     \\
Bulk phase change temperature $T_r$                   & 1620 K \\
Liquidus temperature $\Tliq$                       & 1622.5 K \\
Solidus temperature $\Tsol$                       & 1617.5 K  \\
Latent heat  $L$                &  250800  J/kg    \\
\bottomrule
 \end{tabular}
\end{table}

 \subsubsection{Three-phase melting problem with a volumetric heat source}  \label{sec_3phase_stefan_prob_melting}
We now consider the three-phase version of the melting problem in which the heat flux boundary condition is imposed through a volumetric source term $Q_\text{src}$\footnote{\REVIEW{This benchmark test is provided in IBAMR GitHub within the directory \texttt{examples/phase\_change/ex9}.}}. The position of the heat source remains fixed at $x = 0$, the initial location of the gas-PCM interface. Over time, the heat source strength varies (reduces). The PCM thermophysical properties remain the same as in the two-phase case. 
 
Initially the domain is occupied by gas and solid PCM as illustrated in Fig.~\ref{fig_stefan_melting_dynamics}(B). At $t = 0$, $\Omega^{\text{P}} \in [0, 0.1]\times[0, 0.005]$ and $\Omega^{\text{G}} \in [-0.02, 0] \times [0, 0.005]$. Boundary conditions are set as follows: periodic boundaries in the $y$-direction, zero pressure (outflow) for the flow solver and adiabatic (homogeneous Neumann) for temperature at the left ($x = -0.02$) and right ($x = 0.1$) boundaries. The flux boundary condition $\displaystyle{q" = \left. -\kl \frac{{\rm d} \Tl_\text{exact}}{{\rm d} x} \right|_{x=0}}$ is applied at $x = 0$ via Eq.~\eqref{eq_laser_src}.
 
Gas properties are specified as $\rhog = 0.4$ kg/m$^3$, $\kg = 6.1\times10^{-2}$ W/m$\cdot$K, and $\cpg = 1100$ J/kg$\cdot$K. Viscosity is assumed to be zero in all three phases. A two-level AMR mesh with a uniform coarse cell size of $\Delta x= \Delta y= 0.005/32$ is used for the simulation. Tagging is based on $\REVIEW{\Phi}$ for the PCM-gas interface and on $\grad{\varphi}$ for refining the grid near the mushy zone. A smooth Heaviside function $H$ with $\ncells = 2$ is employed.
 
\begin{figure}
\centering
\includegraphics[width=1.0\linewidth]{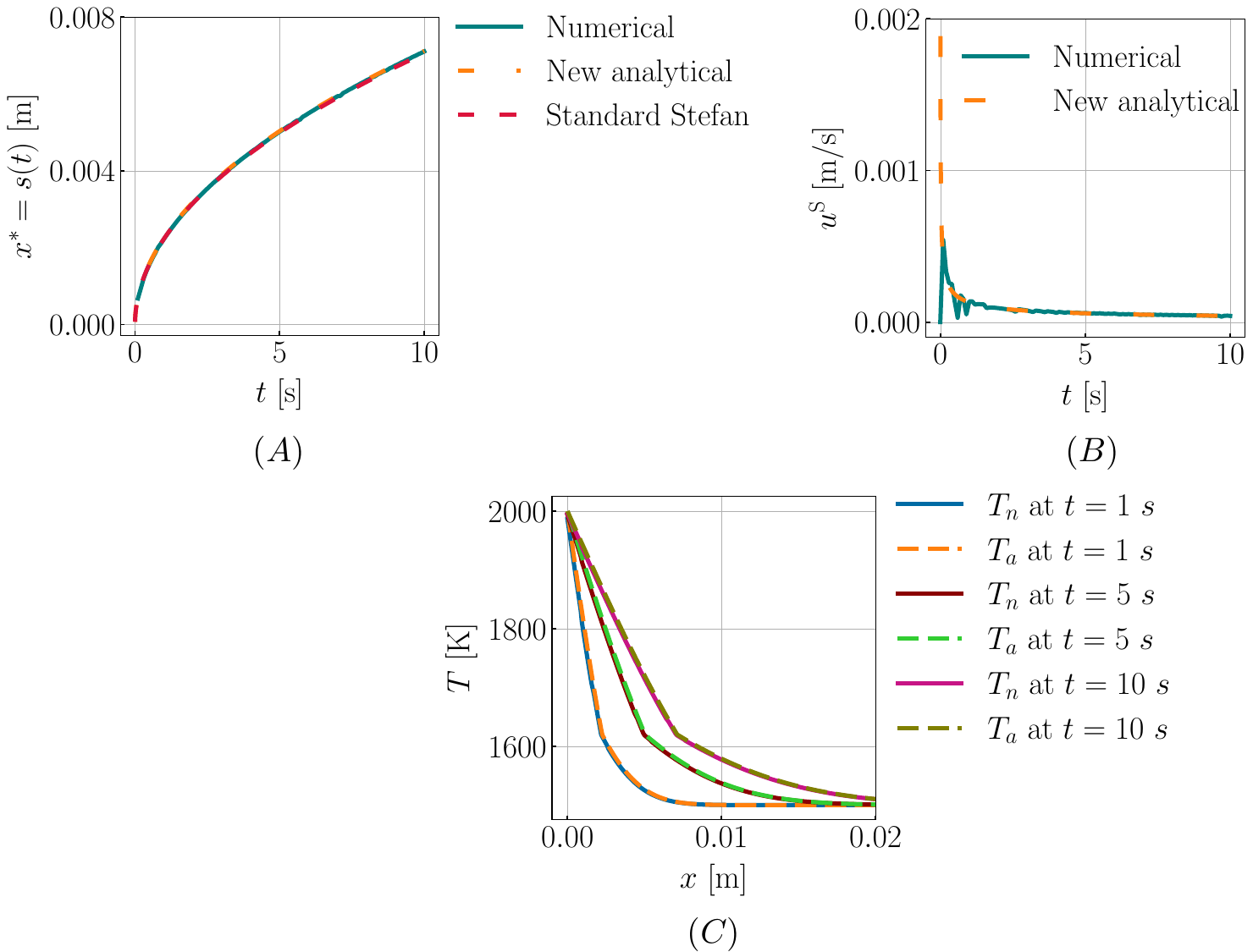}
\caption{Comparison of CFD and analytical solutions for the three phase melting problem with a volumetric heat source term added to the enthalpy equation: (A) interface position; (B) solid velocity; and (C) temperature distribution in the liquid and solid regions.}
\label{fig_stefan_melting_with_gas}
\end{figure}

 Due to the heat source, melting begins and progresses towards the right, as shown in Fig.~\ref{fig_stefan_melting_dynamics}. Comparison between numerical and analytical solutions is presented in Fig.~\ref{fig_stefan_melting_with_gas}. The numerical and analytical solutions are in excellent agreement. \REVIEW{This simple test problem serves as a preliminary check to validate three phase codes aimed at modeling laser-induced PCM melting.}

\subsection{Porosity defect modeling}  \label{sec_bubble_trap}

\begin{figure}
\centering
\includegraphics[width=1.0\linewidth]{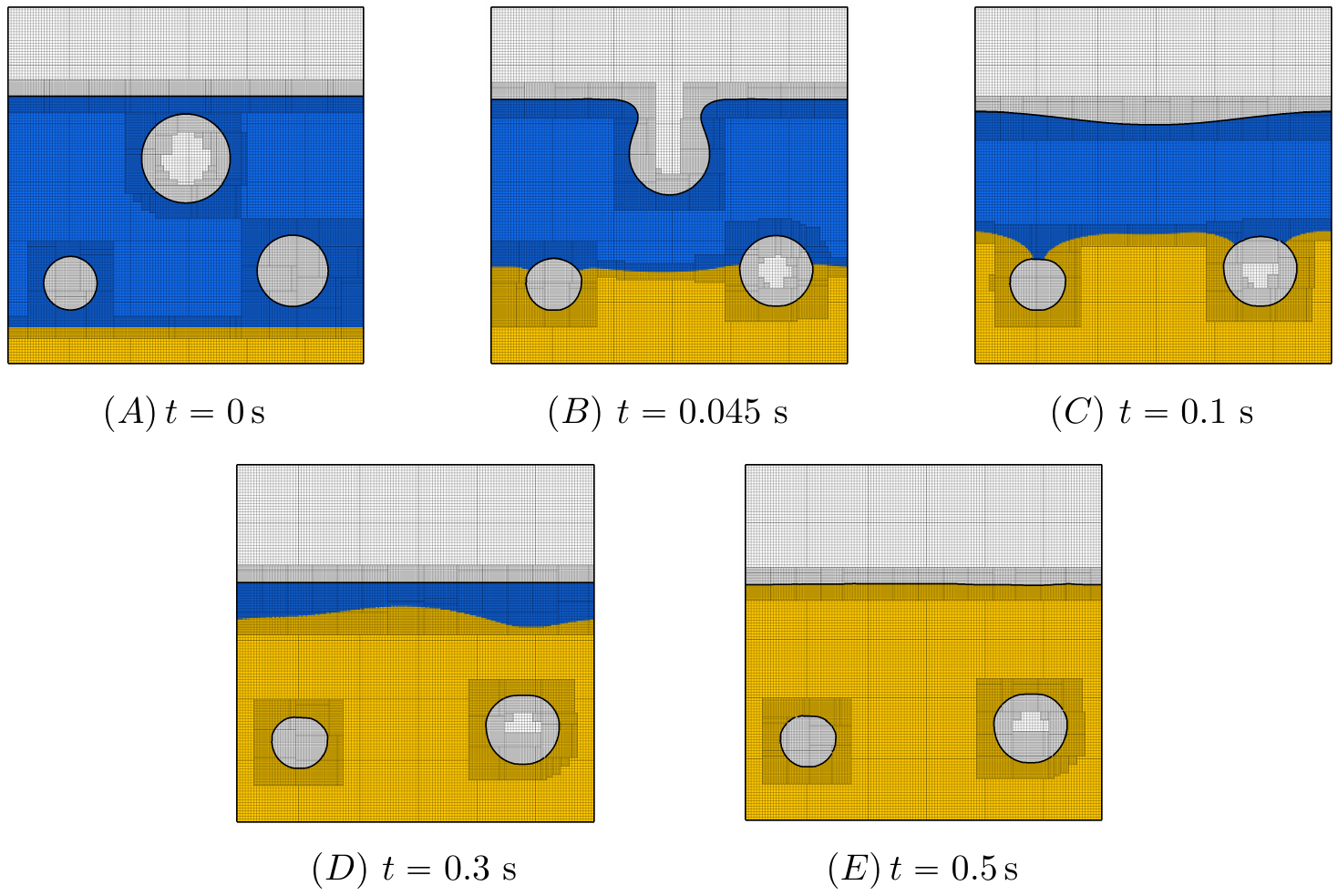}
\caption{Time evolution of metal solidification with three trapped gas bubbles. Solid, liquid and gas regions are depicted in yellow, blue and white colors.}
\label{fig_bubble_trap_dynamics}
\end{figure}

The final test case involves liquid, solid, and gas interactions. The goal is to simulate porosity defects during metal solidification, such as during metal 3D printing and casting. We consider a square computational domain $\Omega \in [0, L]^2$ of length $L = 10^{-2}$ m.~\footnote{\REVIEW{This benchmark test is provided in IBAMR GitHub within the directory \texttt{examples/phase\_change/ex10}.}} Fig.~\ref{fig_bubble_trap_dynamics} depicts the initial phase distribution. The solid phase of the PCM occupies the region up to $y=0.1L$, the liquid phase of the PCM extends above it to $y=0.75L$, and a gas phase occupies the remainder. Three gas bubbles are placed within the liquid phase. The locations and radii of these bubbles can be found in Table~\ref{tab_bubbles_loc}.


Solid and liquid phases of the (aluminum-based) PCM have densities of $\rhol = 2475$ kg/m$^3$ and $\rhos = 2700$ kg/m$^3$, respectively. Viscosity is set to be in the same in the liquid and solid phases $\mul=\mus=1.4\times10^{-3}$ kg/m$\cdot$s. Additional PCM properties are provided in Table~\ref{tab_aluminum_properties}. As for the gas phase, $\rhog = 0.4$ kg/m$^3$, $\kg= 6.1\times10^{-2}$ W/m$\cdot$K, $\cpg = 1100$ J/kg$\cdot$K, and $\mug = 4\times10^{-5}$ kg/m$\cdot$s. The constant surface tension coefficient at the liquid-gas interface is $\sigma = 0.87$ N/m. The gravity force acts in the negative $y$-direction, i.e, $\mathbf{g} = [0, -9.81]$ m/s$^2$.

Boundary conditions are specified as follows: no-slip velocity boundary conditions on the sides and bottom walls, and zero-pressure (outflow) boundary conditions on the top wall. An adiabatic condition is applied to all sides except the bottom wall, where the temperature is set to $T = 0.5T_m$.  $T_m$ denotes the PCM fusion temperature. Initially, all phases have zero velocities. In the solid phase, the initial temperature is $0.5T_m$, while in the liquid and gas phases, it is $1.1T_m$.

The simulation employs a two-level mesh with $N_{x0} \times N_{y0} =   128^2$ coarse level cells, and a refinement ratio of 2 ($\nref=2$). It runs until $t=1$ s with a uniform time step size of $3\times10^{-6}$ s, which is determined by the explicit capillary time step size constraint. The PCM-gas interfacial region is refined using value-based ($|\REVIEW{\Phi}|_{i,j} \leq 2\Delta x_0$) tagging criterion. Gradient-based tagging refines the solid-liquid interfacial regions. As bubbles rise due to buoyancy, solidification occurs concurrently at the bottom wall due to the imposed temperature boundary condition. The movement of bubbles induces convective flow in the liquid below, which helps solidify it more quickly. After a while, the liquid-solid interface entraps the left and right bubbles, while the center bubble breaks off from the liquid-gas interface and enters the gas phase. Due to surface tension, the breaking of the liquid-gas interface initiates capillary waves, which eventually dissipate due to the combined effects of fluid viscosity and freezing. In the meantime, the two remaining bubbles are trapped within the solid. As a result of the slow solidification rate above the bubbles and the motion of the remaining untrapped portion, the left and right bubbles take on the shape of a pancake (flat on the top). Solidification continues until the liquid PCM fully solidifies. This is a shrinkage problem ($R_\rho = \rhos/\rhol > 1$), so the top PCM-gas interface moves downwards due to volume changes, as shown in final Fig.~\ref{fig_bubble_trap_dynamics}(E). Upon completion of solidification, the interfaces remain stationary. At this point, the liquid phase has completely disappeared from the domain. The results of the previous Stefan problem section, as well as this section highlight the strength of our method to (i) handle the appearance and disappearance of phases in the simulation; (ii) properly account for volume changes in the material; and (iii) maintain equilibrium with no spurious phase generation in the absence of a phase change triggering mechanism. 
      
\begin{table}[]
\centering
\caption{Initial location and radii of bubbles.}
\label{tab_bubbles_loc}
\begin{tabular}{clcl}
\midrule
Bubble	&Origin	&Radius \\
\midrule
Left  	&[0.175L, 0.225L]	&0.075L     \\
Center 	&[0.5L, 0.55L]		& 0.125L     \\
Right 	&[0.8L, 0.26L]		&0.1L     \\
\bottomrule
\end{tabular}
\end{table}

\begin{figure}
\centering
\includegraphics[width=0.9\linewidth]{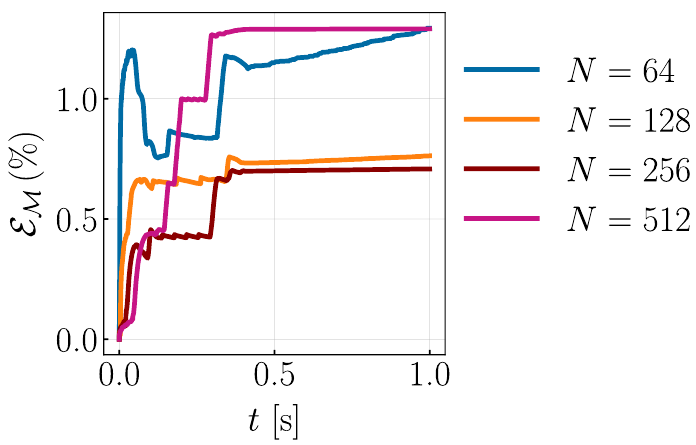}
\caption{\REVIEW{Percentage change in PCM mass $\mathcal{E_M}$ as a function of time for the metal solidification case with three gas bubbles. The coarsest grid $(N_x \times N_y = 64^2)$ uses a uniform time step size of $\Delta t = 3\times10^{-5}$ s. For each successively refined grid, the time step size is reduced by a factor of 3. The temperature interval considered in this case is $ \Tliq - \Tsol = 5$ K.}}
\label{fig_bubble_trap_pcm_mass_3_bubbles}
\end{figure}

\begin{figure}
\centering
\includegraphics[width=0.9\linewidth]{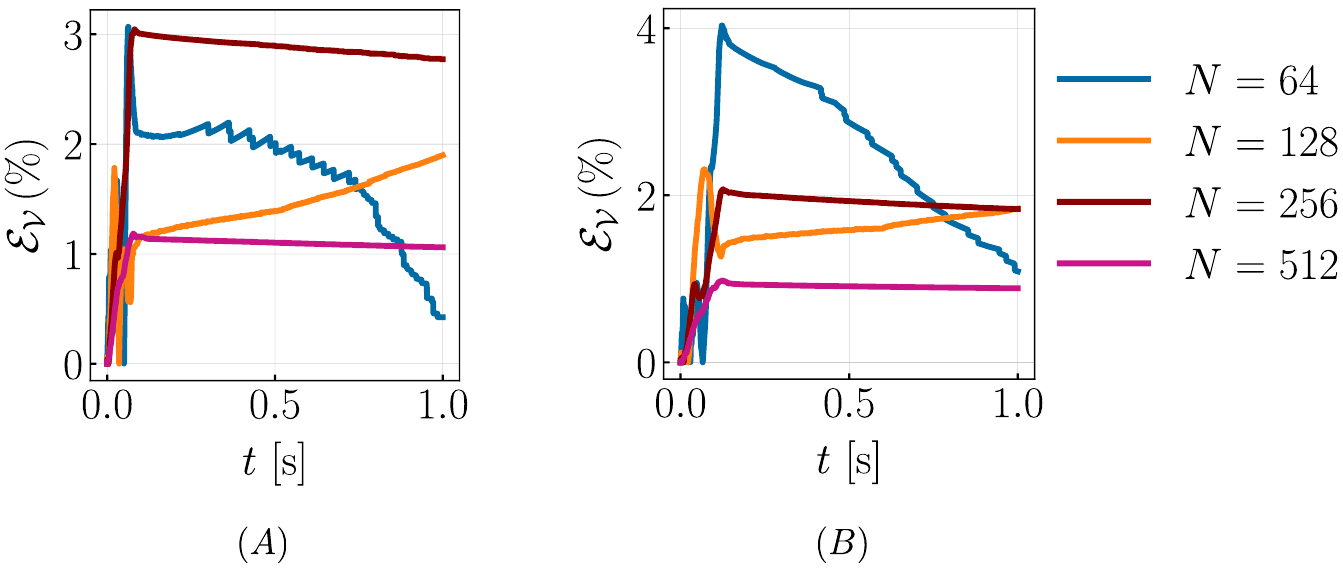}
\caption{\REVIEW{Percentage change in volume $\mathcal{E_V}$ as a function of time for the (A)~left and (B)~right bubbles for the metal solidification case with three gas bubbles. The coarsest grid $(N_x \times N_y = 64^2)$ uses a uniform time step size of $\Delta t = 3\times10^{-5}$ s. For each successively refined grid, the time step size is reduced by a factor of 3. The temperature interval considered in this case is $ \Tliq - \Tsol = 5$ K.}}
\label{fig_bubble_trap_bubble_volume_3_bubbles}
\end{figure}

\begin{figure}
\centering
\includegraphics[width=1.0\linewidth]{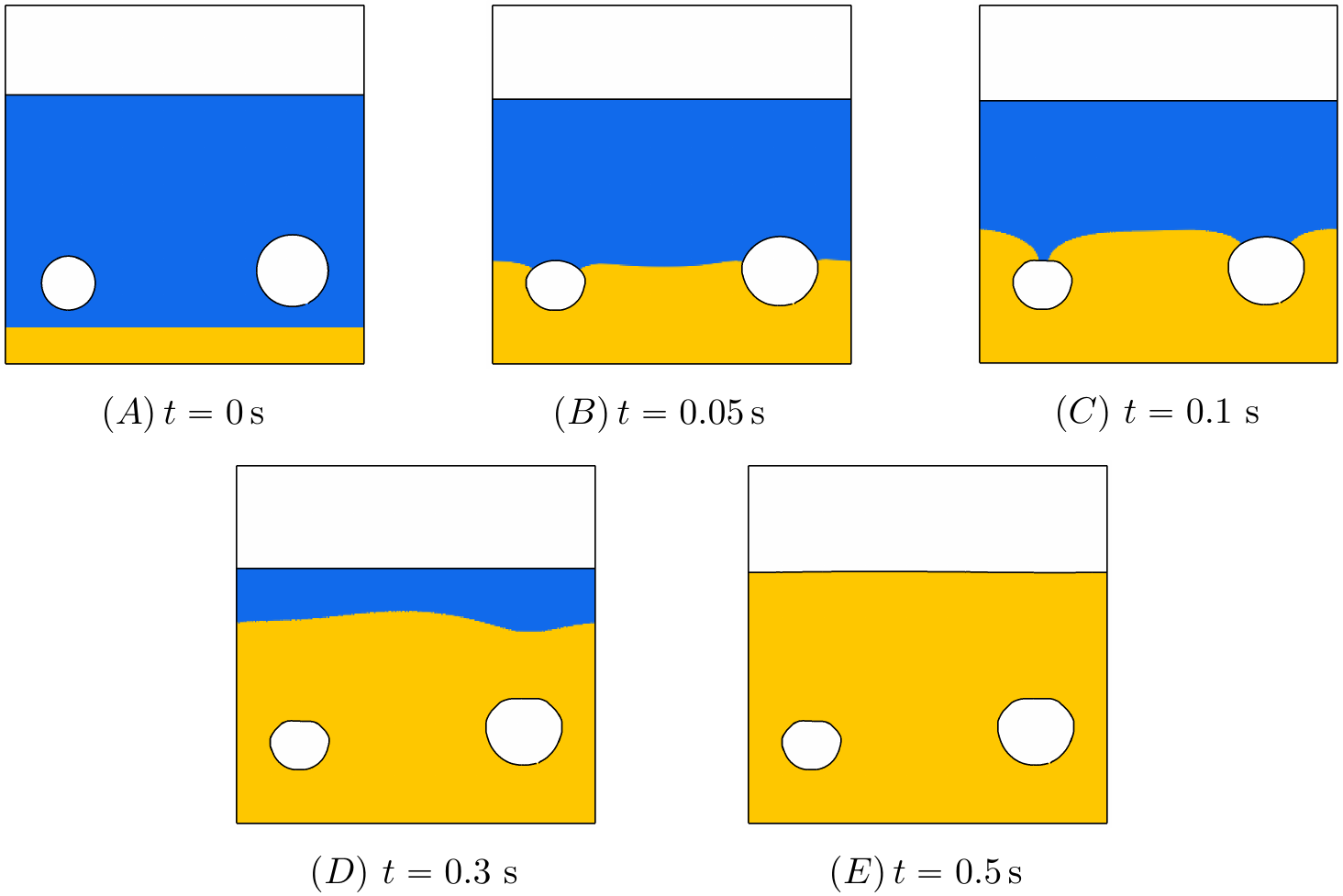}
\caption{Time evolution of metal solidification with two trapped gas bubbles. Solid, liquid and gas regions are depicted in yellow, blue and white colors.}
\label{fig_bubble_trap_two_bubble_dynamics}
\end{figure}

\begin{figure}
\centering
\includegraphics[width=0.9\linewidth]{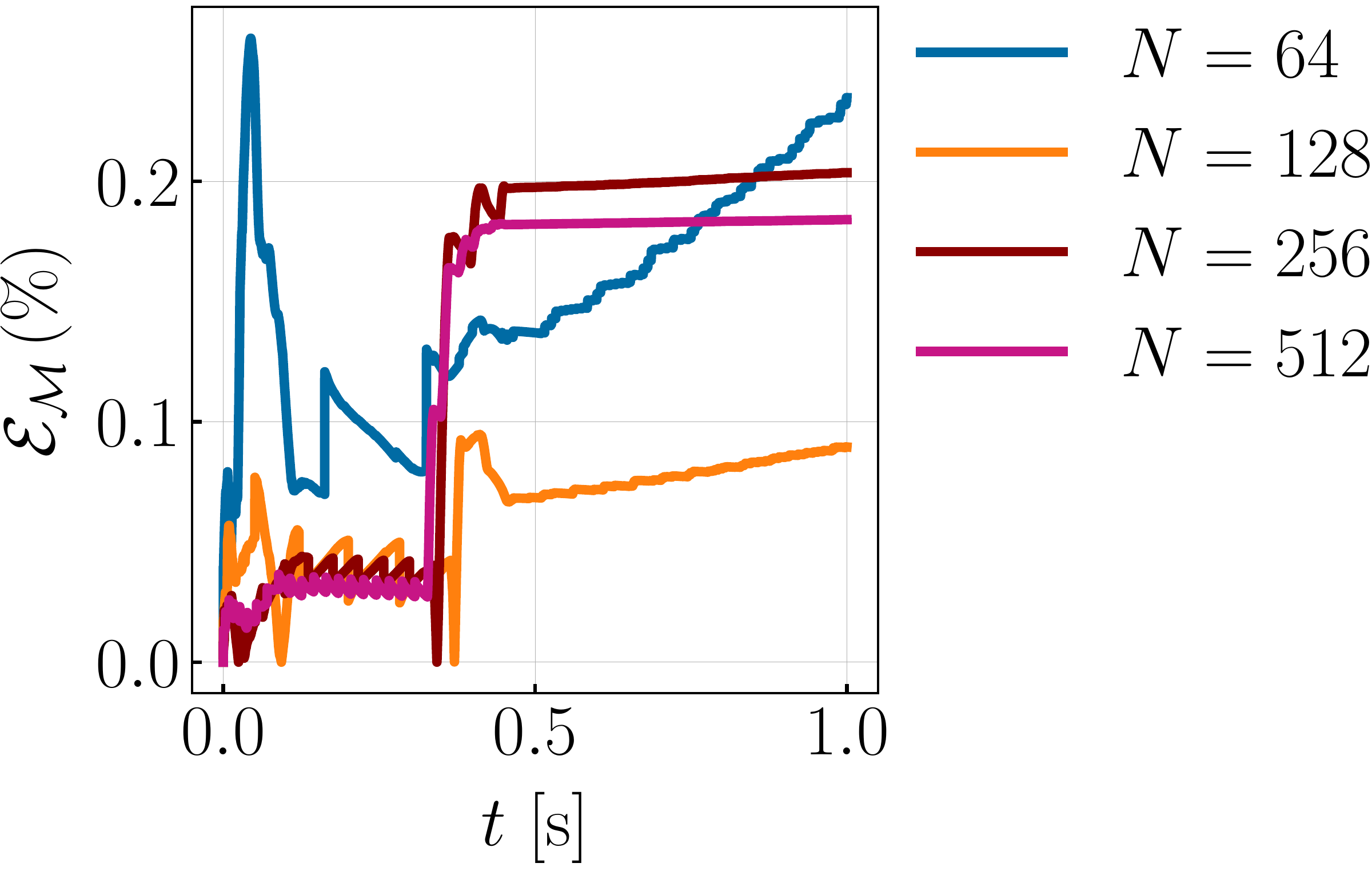}
\caption{Percentage change in PCM mass $\mathcal{E_M}$ as a function of time for the metal solidification case with two trapped gas bubbles. The coarsest grid $(N_x \times N_y = 64^2)$ uses a uniform time step size of $\Delta t = 3\times10^{-5}$ s. For each successively refined grid, the time step size is reduced by a factor of 3. The temperature interval considered in this case is $ \Tliq - \Tsol = 5$ K.}
\label{fig_bubble_trap_pcm_mass}
\end{figure}

\begin{figure}
\centering
\includegraphics[width=0.9\linewidth]{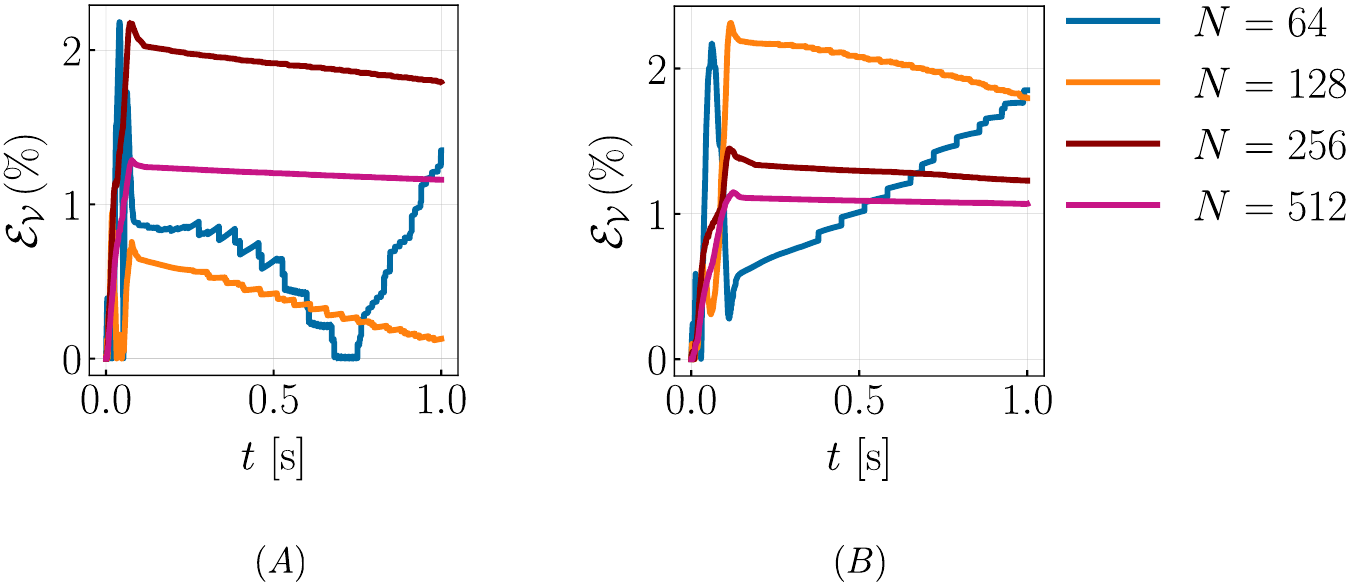}
\caption{\REVIEW{Percentage change in volume $\mathcal{E_V}$ as a function of time for the (A)~left and (B)~right bubbles for the metal solidification case with two trapped gas bubbles. The coarsest grid $(N_x \times N_y = 64^2)$ uses a uniform time step size of $\Delta t = 3\times10^{-5}$ s. For each successively refined grid, the time step size is reduced by a factor of 3. The temperature interval considered in this case is $ \Tliq - \Tsol = 5$ K.}}
\label{fig_bubble_trap_bubble_volume}
\end{figure}

\REVIEW{Furthermore, we quantify the mass/volume conservation of our scheme in this case. The PCM mass is computed numerically as
\begin{equation}
\mathcal{M}_{\rm PCM}(t) = \int_\Omega \left[ \rhol (H\varphi) + \rhos (H-H\varphi) \right] \, \d V.
\end{equation}
The relative change in the PCM mass is obtained using Eq.~\eqref{eq_rel_change} and plotted against time for four uniform grids of size $N_x \times N_y = \{64^2, 128^2, 256^2, 512^2\}$ in Fig.~\ref{fig_bubble_trap_pcm_mass_3_bubbles}. The coarsest grid $(N_x \times N_y = 64^2)$ uses a uniform time step size of $\Delta t = 3 \times 10^{-5}$ s. For each successively refined grid, the time step size is reduced by a factor of 3, respecting the explicit capillary time step size restriction. 

It can be observed that there are two stages where PCM mass error increases over time. Initially, the PCM mass error increases rapidly until the center bubble is completely released into the air. Afterward, the PCM mass error remains constant for some time before growing again when the PCM-gas interface and liquid-solid interfaces approach each other. Around $t = 0.3$ s, the gas-PCM and liquid-solid interfaces come close, resulting in an error in mass calculation because two different diffuse interfaces meet and equilibrate rapidly, causing the PCM to solidify completely. The non-linear convergence rate is attributed to errors associated with interface breaking and the interaction of two diffuse interfaces. The relative change in bubble volume is illustrated in Fig.~\ref{fig_bubble_trap_bubble_volume_3_bubbles}. For all grids considered, the maximum volume change for the two bubbles is 4\%. Given that the level set method leads to mass/volume loss, especially at coarse grid resolutions, this is an acceptable level of volume change.

We rerun the simulation after removing the center bubble from the domain to obtain meaningful convergence rates~\footnote{Convergence rates are well-defined for smooth functions/dynamics.}. This prevents interface breaking and ensures smooth dynamics.  Fig.~\ref{fig_bubble_trap_two_bubble_dynamics} shows the evolution of the interfaces. Solidification and bubble trapping dynamics are very similar to the three-bubble case discussed above. The same uniform grids are used to study mass errors as in the three bubble setup.  The relative change in PCM mass is plotted against time in Fig.~\ref{fig_bubble_trap_pcm_mass}.  It can be observed that as the grid cell size $\Delta$  decreases, the error in PCM mass, $\mathcal{E_M}$, also decreases until $t = 0.3$ s. In the three bubble case, interface breaking causes PCM mass change errors to exhibit non-uniform convergence rates. In the two-bubble case without interface breaking, PCM mass change errors converge under grid refinement. Around $t = 0.3$ s, the interface between gas-PCM and liquid-solid comes close to one another, leading to an increase in mass errors, which is similar to the three bubble setup. However, the \% mass change of the PCM for all grids is quite low ($\approx 0.2\%$) compared to the bubble breaking case (roughly 1.4\%). The relative change in bubble volume is illustrated in Fig.~\ref{fig_bubble_trap_bubble_volume}. For all grids considered, the maximum volume change for the two bubbles is 2\% compared to 4\% when bubble breaking is considered. }

\section{Conclusions and discussion}

This work extends our previously introduced low Mach enthalpy method for simulating phase change materials (PCMs) interacting with an ambient gas phase. We implemented consistent time integration schemes for the mass, momentum, and enthalpy equations to improve method's stability. Incorporating an additional stabilizing term into the momentum and enthalpy equations further enhances stability. This is even when using different time integrators for mass and momentum (or mass and enthalpy) equations. The scheme's stability was validated by simulating the advection of an  \REVIEW{isothermal droplet} in inviscid flow and the thermocapillary migration of gas bubbles within a liquid PCM. To reduce the computational cost of simulations, we propose a gradient-based tagging method for adaptively refined grids. Its effectiveness was demonstrated using a two-phase Stefan problem with density change. Additionally, we introduced a simple analytical model for validating CFD simulations involving heat sources (like lasers) that drive phase change phenomena. We also demonstrated our method's ability to simulate porosity defects in metal AM and casting by modeling a three-phase liquid-solid-gas interaction problem. Furthermore, a three-phase surface tension model based on a PDE-based field extension technique was developed. Future work will focus on incorporating an evaporation and condensation model into the low Mach enthalpy approach. This will enable simulations of four-phase flows, including melting, solidification, evaporation, and condensation occuring simultaneously.

\section*{Acknowledgements}
R.T. and A.P.S.B~acknowledge support from NSF awards OAC 1931368 and CBET CAREER 2234387. Compute time on SDSU's high performance computing cluster Fermi is greatly acknowledged.


\section*{Supplementary material}
\noindent \textbf{Supplementary video S1}: Isothermal advection of a \REVIEW{droplet} using AMR grids (see Sec.~\ref{sec_bubble advection_enthalpy}). \\
\noindent \textbf{Supplementary video S2}: Time evolution of the liquid-solid interface and AMR mesh (generated using $\grad{\varphi}$-based tagging) for the three-phase melting problem with a volumetric heat source (see Sec.~\ref{sec_3phase_stefan_prob_melting}).
\noindent \textbf{Supplementary Video S3}: Time evolution of the solid, liquid, and gas domains during metal solidification with trapped gas bubbles demonstrating porosity defects (see Sec.~\ref{sec_bubble_trap}).


\appendix
\section{Analytical solution of the two-phase Stefan problem with volume changes} \label{sec_2phase_analytical}

We derive the analytical solution to the two-phase Stefan problem discussed in Sec.~\ref{sec_stefan_prob_melting}. The PCM is considered as a one-dimensional slab $\Omega := 0 \le x \le l$,  with the solid phase occupying the entire domain at $t=0$. In the beginning, the temperature is uniform throughout the domain and is less than the melting temperature ($T_i < T_m$). On the right end, the boundary is open, while on the left, it is closed. Imposing a temperature of $T_o > T_m$ at the left boundary melts the PCM. Solid phase thermophysical properties are $\rhos$, $\cps$, and $\ks$, while liquid phase thermophysical properties are $\rhol$, $\cpl$, and $\kl$. In each phase, these properties are assumed to be constant. When the PCM melts, the melt front $x^* = s(t)$ moves rightward. The solid-to-liquid density ratio is denoted as  $R_\rho = \rhos/\rhol$. When $R_\rho > 1$, the liquid expands upon melting; when $R_\rho < 1$, the liquid shrinks upon melting. In~\cite{thirumalaisamy2023lowmach}, we presented an analytical solution to the two-phase Stefan solidification problem. A very similar approach is followed here, with some minor changes to account for PCM melting instead of solidification. Our previous work~\cite{thirumalaisamy2023lowmach} provides additional details about the derivation, including jump conditions across the phase-change interface.

The governing equations in the solid and liquid phases are
\begin{subequations} \label{eq_nstd_sp}
\begin{alignat}{2}
\rhos \cps \left( \frac{\partial \Ts}{\partial t} + \us \frac{\partial \Ts}{\partial x} \right) &=\ks\frac{\partial^2 \Ts}{\partial x^2} \, \in  \Omegas(t),  \label{eq_temp_solid}  \\
\rhol \cpl \left( \frac{\partial \Tl}{\partial t} + \ul \frac{\partial \Tl}{\partial x} \right)&=\kl\frac{\partial^2 \Tl}{\partial x^2} \,  \in \Omegal (t), \label{eq_temp_liquid}
\end{alignat}
\end{subequations}
in which $\us$ and $\ul$ are the velocities of the solid and liquid phases, respectively. The melt front moves with a velocity $u^* = \d{s}/\d{t}$. The velocity in the liquid domain $\Omegal(t) := 0 \le x < s(t)$ is zero, i.e., $\ul \equiv 0$. This follows from the 1D continuity equation $\partial \ul/ \partial x= 0$ and the no-slip condition at the left boundary $x = 0$. The jump in velocity across the interface~\cite{thirumalaisamy2023lowmach} gives 
\begin{equation}
\llbracket u \rrbracket =  \us (s^{+},t )- \ul(s^{-}, t) =  \us (s^{+},t ) = \left(1-\frac{1}{R_\rho}\right) \dd{s}{t}, \label{eqn_vel_jump}
\end{equation}
in which $s^{+}$ and $s^{-}$ represent spatial locations just ahead and behind the interface. The continuity equation in the solid domain $\partial \us/ \partial x= 0$, yields a constant velocity in the solid domain $\Omegas(t) := s(t) < x \le l$  as $\us(\Omegas,t) \equiv  \us (s^{+},t ) = \left(1-\frac{1}{R_\rho}\right) \dd{s}{t}$. Two boundary conditions $T(x= 0,t) = T_0$ and $T(x= l,t) = T_i$, and three interfacial conditions
\begin{align}
&\Ts(x^*, t) = \Tl(x^*, t) = T_m, \label{eq_interface_temp}\\
&\rhol\left[(\cpl - \cps)(T_m - T_r) + L+\frac{1}{2}\left(1-\frac{1}{R_\rho^2}\right) \left( \dd{s}{t} \right)^2\right] \dd{s}{t}    \nonumber \\
& = \left( \ks\frac{\partial \Ts}{\partial x}  -\kl \frac{\partial \Tl}{\partial x}  \right)_{x^*}.  \label{eq_stefan_condition}
\end{align}
allow us to find the temperature distribution in the liquid ($\Tl(x,t)$) and solid ($\Ts(x,t)$) domains, as well as the interface position $s(t) = 2\sqrt{\alphas t} \lambda(t)$.  The interface condition Eq.~\eqref{eq_stefan_condition} is known as the Stefan condition. The Stefan condition for the present melting problem differs slightly from that provided for the solidification problem in our previous work (Eq.~12 of \cite{thirumalaisamy2023lowmach}). The Stefan condition for melting and solidification problems becomes the same when $R_\rho = 1$, which is typically the case considered in textbooks and research papers. 

Eqs.~\eqref{eq_temp_solid} and \eqref{eq_temp_liquid} admit similarity solutions of the form~\cite{thirumalaisamy2023lowmach}
\begin{align}
\Ts(x,t) &= T_i + A(\lambda(t))\,\text{erfc}\left(\frac{x}{2\sqrt{\alphas t}} - \frac{s(t)}{2\sqrt{\alphas t}}\left(1-\frac{1}{R_\rho}\right)\right), \label{eq_Tprofile_solid} \\
\Tl(x,t) &=  T_o + B(\lambda(t))\,\text{erf}\left(\frac{x}{2\sqrt{\alphal t}}\right). \label{eq_Tprofile_liquid}
\end{align}
Here, $\alphas = \ks/(\rhos \cps)$, $\alphal = \kl/(\rhol \cpl)$ are the solid and liquid thermal diffusivities, respectively, and $\lambda(t) = \frac{s(t)}{2\sqrt{\alphas t}}$ is a yet to be determined function of time. The coefficients $A(\lambda(t))$ and $B(\lambda(t))$ are found using the interface temperature condition as
\begin{equation*}
T_i + A(\lambda(t))\,\text{erfc}\left( \frac{\lambda(t)}{R_\rho} \right) = T_o + B(\lambda(t))\,\text{erf}\left(\frac{s(t)}{2\sqrt{\alphal t}}\right) = T_m.
\end{equation*}
This yields
\begin{equation}
A(\lambda(t)) = \frac{T_m-T_i}{\text{erfc}\left( \frac{\lambda(t)}{R_\rho} \right) } \qquad \text{and}   \qquad B(\lambda(t)) =  \frac{T_m-T_o}{\text{erf}\left(\lambda(t) \sqrt{\frac{\alphas}{\alphal}} \right)}.
\end{equation}
Thus, the temperature distribution in the solid phase is given by
\begin{equation}
\Ts(x,t) = T_i + (T_m-T_i)  \frac{\text{erfc}\left(\frac{x}{2\sqrt{\alphas t}} - \lambda(t) \left(1-\frac{1}{R_\rho}\right)\right)} {\text{erfc} \left( \frac{\lambda(t)}{R_\rho}\right)}
\label{eq_temp_solid2}
\end{equation}
and in the liquid phase by
\begin{equation}
\Tl(x,t) = T_o + (T_m-T_o) \frac{\text{erf}\left(\frac{x}{2\sqrt{\alphal t}}\right)}{\text{erf}\left(\lambda(t) \sqrt{\frac{\alphas}{\alphal}} \right)}.
\label{eq_temp_liquid2}
\end{equation}

Substituting the liquid and solid temperature profiles, along with the interface position $s(t) = 2\lambda(t)\sqrt{\alphas t}$ into the Stefan condition Eq.~\eqref{eq_stefan_condition}, an equation for $\lambda(t)$ is obtained 
\begin{align}
& \rhol \left[ L^{\rm eff} + \frac{\left(1-\frac{1}{R_\rho^2}\right)}{2}\left(\frac{\lambda^2 \alphas}{t}\right)\right]\lambda \sqrt{\alphas} =  \nonumber \\
& -\ks \frac{T_m-T_i} {\text{erfc}\left(\frac{\lambda}{R_\rho}  \right)} \frac{e^{-\left(\frac{\lambda}{R_\rho}\right)^2}  }{\sqrt{\pi \alphas}}- \kl\frac{T_m-T_o}{\text{erf}\left(\lambda \sqrt{\frac{\alphas}{\alphal}} \right)} \frac{e^{-\lambda^2 \alphas / \alphal}}{\sqrt{\pi \alphal}}.
\label{eq_lambda_differential}
\end{align}
Here, $L^{\rm eff} = L + (\cpl - \cps)(T_m - T_r)$ represents effective latent heat. In this equation, only the leading-order $1/\sqrt{t}$ term is retained, while other terms related to the derivative of $\lambda(t)$ are dropped (see \cite{thirumalaisamy2023lowmach} for details). 


\section*{Bibliography}
\begin{flushleft}
 \bibliography{laser_bibliography.bib}
\end{flushleft}

\end{document}